\documentclass[]{article}

\usepackage[margin=1in]{geometry}
\usepackage[numbers,sort&compress]{natbib}
\usepackage{enumitem}
\newcommand{\volumeheader}[3]{}
\newcommand{\upstairs}[1]{\textsuperscript{#1}}
\newcommand{\affilone}{1}
\newcommand{\copyrightnotice}{}

\RequirePackage{amsthm,amsmath,amsfonts,amssymb}
\usepackage{hyperref}       
\usepackage{url}            
\usepackage{booktabs}       
\usepackage{nicefrac}       
\usepackage{microtype}
\usepackage{graphicx}
\usepackage{subfig}
\usepackage{bm}
\usepackage{xcolor}
\usepackage{lineno}
\usepackage{tikz}
\usetikzlibrary{patterns}
\usepackage{xcolor}
\usepackage{multirow}
\usepackage{algorithm}
\usepackage{algorithmic}
\usepackage{soul}
\usepackage{tabularx}
\usepackage{makecell}

\newcolumntype{Y}{>{\raggedright\arraybackslash}X}
\newcolumntype{P}[1]{>{\raggedright\arraybackslash}p{#1}}

\newcommand\cB{\mathcal{B}}

\newcommand\cO{\mathcal{O}}

\newcommand\cM{\mathcal{M}}

\newcommand\bV{\mathbb{V}}
\newcommand\nd{\textnormal{d}}

\newcommand{\bR}{\mathbb{R}}
\newcommand{\reach}{\rm{reach}}
\newcommand{\dhaus}{\nd_{\textnormal{H}}}

\theoremstyle{plain}
\newtheorem{theorem}{Theorem}[section]

\newtheorem{lemma}[theorem]{Lemma}

\newtheorem*{remark}{Remark}
\newtheorem{definition}[theorem]{Definition}

\graphicspath{{figs/}{../figs/}}

\begin{document}


\volumeheader{0}{0}{00.000}

\begin{center}

  {\LARGE Manifold Fitting:  A Review of Methods and Applications\par}

  \thispagestyle{empty}
  
  \vspace*{0.75in}

  \begin{tabular}{cc}
    Zhigang Yao\upstairs{\affilone,*}, Jiaji Su\upstairs{\affilone}
   \\[0.25ex]
   {\small \upstairs{\affilone} National University of Singapore} \\
  \end{tabular}
  
  \begingroup
  \renewcommand{\thefootnote}{*}
  \footnotetext{Corresponding author: \href{mailto:zhigang.yao@nus.edu.sg}{zhigang.yao@nus.edu.sg}.}
  \endgroup

  \vspace*{0.4in}

\begin{abstract}
With data growing in scale and complexity, traditional linear dimension reduction techniques are becoming inadequate in some settings. Manifold fitting offers an important alternative by capturing low-dimensional latent geometric structures within high-dimensional spaces. This capability allows it to support downstream analysis in complex data settings.

In this review, we explore the development and applications of manifold fitting. First, we introduce the basic concepts of manifold fitting and distinguish it from related techniques such as manifold embedding and denoising. We review the development of manifold fitting with three distinct stages: early nonparametric statistical methods, insights from mathematical analysis, and contemporary practical statistical approaches. 

Furthermore, we present diverse applications of manifold fitting, particularly in neural networks and bioinformatics, which illustrate its utility in complex data scenarios. Despite considerable progress, manifold fitting remains a fertile area for research. Many theoretical and practical questions remain unanswered, and ongoing investigations will further clarify its role in modern data science as a geometric tool for a wide range of data analysis challenges.
\end{abstract}
\end{center}

\vspace*{0.15in}
\hspace{10pt}
  
\copyrightnotice

\section*{Media Summary}
Researchers have developed a powerful new technique called manifold fitting that is revolutionizing the way we analyze and understand complex high-dimensional data. This method improves on traditional linear-based techniques by offering more precise ways to reduce and simplify large data sets into understandable forms. Manifold fitting is particularly useful in fields with low signal-to-noise ratio, where efficient management of vast amounts of data can lead to faster progress and deeper insights.

This review traces the evolution of this technique from its early statistical roots to its modern applications, which now incorporate sophisticated mathematical tools and practical methods. Examples show how manifold fitting is already helping scientists better understand the underlying structure of images and genetic data from individual cells.

Despite its success, manifold fitting still has enormous untapped potential. The researchers encourage continued exploration to fully integrate and expand this tool into data science. Their work points to a future where manifold fitting could become a standard tool, aiding countless scientific and technological endeavors by making complex data more accessible and useful.

\section{Introduction}
Over the last few decades, data dimensionality has increased significantly. Traditionally, to manage such high-dimensional data, methods like Principal Component Analysis (PCA) have been commonly used. PCA simplifies data representation by projecting data points onto a lower-dimensional linear subspace through the computation of eigenvectors from the sample covariance matrix. This process facilitates easier visualization and analysis. However, linear methods like PCA are limited to capturing only linear relationships and often fail to adequately represent complex, nonlinear patterns. To address these limitations, more advanced nonlinear techniques are necessary, particularly those involving the concept of a manifold.

In mathematics, a manifold is a topological space that resembles Euclidean space locally around each point. The concept of manifold is fundamental to many areas of geometry and modern mathematical physics, as it describes complex structures through the well-understood properties of simpler spaces. In high-dimensional data analysis, this idea appears through the manifold hypothesis. Interpreted carefully, the hypothesis is not that all data sets lie exactly on a low-dimensional manifold, but rather that in some problems the data-generating distribution may be well approximated by observations supported on, or concentrated near, a class of smooth low-dimensional manifolds with controlled geometric complexity. Under this view, the key claim is that the ambient dimension can be much larger than the intrinsic dimension needed to describe the dominant variability of the data. Such a perspective is plausible when a system is governed by relatively few latent degrees of freedom, so that its dominant variability is effectively low-dimensional even though it is recorded in a high-dimensional form, but it should be understood as a problem-dependent modeling assumption rather than a universal fact. Classical embedding results, such as the Whitney and Nash embedding theorems, provide background for representing smooth manifolds in Euclidean space. Throughout this review, we restrict attention to smooth embedded submanifolds of Euclidean space, excluding immersed or self-intersecting objects; orientability is not assumed unless explicitly needed.

This statistical viewpoint is made particularly explicit by \citet{fefferman2016testing}. In plain terms, one may ask whether the observed distribution is well approximated by some manifold in an admissible class characterized by intrinsic quantities such as dimension, reach, and overall size or volume. At a high level, there are then two possible answers: either there exists such a manifold that yields small approximation error, or no manifold in the admissible class achieves comparably small error. Framed in this way, the manifold hypothesis becomes a question of model adequacy, which naturally leads to the basic statistical problem of how many samples are needed to decide, with high confidence, whether a low-dimensional manifold model provides a satisfactory explanation of the data.

Once the question is posed in this way, a natural next step is to move from testing adequacy to explicitly constructing a manifold that explains the data well. This leads directly to manifold fitting, where the goal is to estimate a smooth manifold object in the ambient space from sampled observations, typically by minimizing or approximately minimizing a suitable distance-based loss. From this perspective, manifold fitting is a central paradigm within the broader manifold-learning landscape, complementing methods that instead seek either a low-dimensional representation in another space or a pointwise correction of noisy observations. As suggested by \citet{yao2023manifold}, based on the overall target, these techniques can be broadly classified into three categories: manifold embedding, manifold denoising, and manifold fitting.

To orient the reader before the detailed review, Table~\ref{tab:manifold-paradigms} places manifold fitting within a broader manifold-learning landscape and also includes a neighboring line of PCA-type nonlinear extensions. It highlights how these related directions differ in target, output, and relation to the latent-manifold-plus-noise viewpoint.

\begin{table}[htbp]
    \centering
    \caption{Compact comparison of related manifold-learning paradigms and neighboring PCA-type nonlinear extensions. The representative references are included only to highlight a few characteristic contributions in each direction and are not meant to be exhaustive.\\[-5pt]}
    \label{tab:manifold-paradigms}
    \footnotesize
    \setlength{\tabcolsep}{2pt}
    \renewcommand{\arraystretch}{1.5}
    \renewcommand{\tabularxcolumn}[1]{>{\raggedright\arraybackslash}m{#1}}
    \begin{tabularx}{\textwidth}{@{}>{\raggedright\arraybackslash}m{2.2cm}
                                    >{\raggedright\arraybackslash}m{2.1cm}
                                    >{\raggedright\arraybackslash}m{2.5cm}
                                    >{\raggedright\arraybackslash}X
                                    >{\raggedright\arraybackslash}m{3.3cm}@{}}
        \toprule
        Paradigm & Goal & Output & Relation to Eq.\,\eqref{eq:Add_model} & Representative references \\
        \midrule
        Hypothesis testing
        & Model adequacy
        & Test; admissible class
        & Explicit latent-manifold model class
        & \citet{fefferman2016testing} \\

        Embedding
        & Low-dimensional representation
        & Coordinates; embedding map
        & Usually not explicit ambient-space recovery
        & \makecell[l]{\citet{tenenbaum2000global}\\ \citet{roweis2000nonlinear}\\
        \citet{zhang2004principal} }\\

        Regularization
        & Geometry-aware learning
        & Regularized predictor; classifier
        & Uses manifold structure as inductive bias; not explicit manifold estimation
        & \makecell[l]{\citet{belkin2006manifold}\\ \citet{meilua2024manifold}} \\

        Diffusion geometry
        & Diffusion-based geometric analysis
        & Diffusion coordinates; operator
        & May encode intrinsic geometry without explicit recovery under Eq.\,\eqref{eq:Add_model}
        & \makecell[l]{\citet{coifman2006diffusion}\\
        \citet{nadler2006diffusion} \\
        \citet{singer2006graph}
        } \\

        Denoising
        & Noise reduction near a manifold
        & Denoised points; local summaries
        & \makecell[l]{Directly aligned with latent \\manifold plus noise;\\mainly pointwise recovery}
        & \makecell[l]{\citet{wang2010manifold}\\ \citet{sober2020manifold}\\ \citet{luo2020differentiable}} \\

        Fitting
        & Ambient-space manifold estimation
        & Manifold estimator $\widehat{\mathcal{M}}$
        & Direct latent manifold plus noise recovery
        & \makecell[l]{\citet{genovese2012manifold}\\
        \citet{mohammed2017manifold}\\
        \citet{yao2019manifold}\\
        \citet{yao2023manifold}} \\
        
        Nonlinear / intrinsic PCA
        & PCA-type summary of dominant variation
        & Principal curve / flow; geodesic principal components or subspaces
        & Typically assumes the ambient or intrinsic space is already given and seeks PCA-like summaries of variation
        & \makecell[l]{\citet{hastie1989principal}\\
        \citet{huckemann2010intrinsic}\\
        \citet{principal_nested_spheres}\\
        \citet{panaretos2014principal}\\
        \citet{yao2024principalsubmanifolds}\\
        \citet{su2025principal}
        } \\
        \bottomrule
    \end{tabularx}
\end{table}

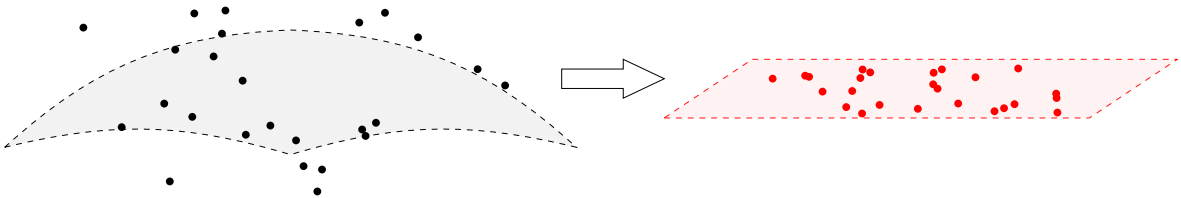
\begin{figure}[htbp]
    \centering
    \resizebox{0.95\textwidth}{!}{
\begin{tikzpicture}[x=0.75pt,y=0.75pt,yscale=-1,xscale=1]
    \clip (0, 0) rectangle (800, 150);
    \pgfmathsetseed{2024}
    
    \fill[gray!10] (0,110) to[bend right=20] (195,30) 
                   to[bend right=20] (390,110)
                   to[bend left=15] (195,115)
                   to[bend left=15] (0,110);
                   
    \draw [dashed] (0,110) to[bend right=20] (195,30) 
                   to[bend right=20] (390,110)
                   to[bend left=15] (195,115)
                   to[bend left=15] (0,110);
    \foreach \i in {1,...,25} {
      \pgfmathsetmacro{\x}{rand*150 + 195}
      \pgfmathsetmacro{\y}{rand*70 + 80}
      \fill (\x, \y) circle (2pt);
      }

    \draw[shift={(380,50)}] (0,6.63) -- (42,6.63) -- (42,0) -- (70,13.25) -- (42,26.5) -- (42,19.88) -- (0,19.88) -- cycle ;
    \draw[shift={(450,00)}, dashed, red] (0,90) -- ++(290,0) -- ++(60,-40) -- ++(-290,0) -- cycle;
    \fill[shift={(450,00)}, pink, opacity=0.2] (0,90) -- ++(290,0) -- ++(60,-40) -- ++(-290,0) -- cycle;
    
    \foreach \i in {1,...,25} {
      \pgfmathsetmacro{\x}{rand*110 + 625}
      \pgfmathsetmacro{\y}{rand*18 + 70}
      \fill [red] (\x, \y) circle (2pt);
      }
\end{tikzpicture}
}
    \caption{Manifold Embedding: This illustration shows observed data points (black) distributed around a latent manifold (gray surface). Manifold embedding seeks a low-dimensional representation in another space (light red parallelogram), producing embedded points (red points) intended to reflect intrinsic-coordinate-like information or other geometric or spectral structure.}
    \label{fig:Illus-embedding}
\end{figure}

\noindent
\textit{Manifold embedding}, a technique for discovering low-dimensional representations of high-dimensional data sets that lie close to unknown low-dimensional manifolds, has been central to the development of dimensionality reduction, visualization and clustering methods since the early 21st century. Rather than pursuing a single notion of fidelity, manifold embedding methods are designed to preserve different structures of the data. Depending on the method, the target may be geodesic or intrinsic-distance-motivated structure, local linear or local tangent structure, neighborhood structure, or spectral structure associated with graph Laplacians and the Laplace--Beltrami operator. The output is therefore a low-dimensional representation in another space that is useful for visualization, clustering, or subsequent analysis, but it is not itself an estimated manifold in the original ambient space.

Extensive research has been conducted on manifold embedding algorithms such as Isomap \citep{tenenbaum2000global}, Locally Linear Embedding \citep{roweis2000nonlinear}, Hessian Eigenmaps \citep{donoho-hessian-lle}, Laplacian Eigenmaps \citep{belkin2003laplacian}, Diffusion Maps \citep{coifman2006diffusion}, Local Tangent Space Alignment \citep{zhang2004principal}, and Uniform Manifold Approximation and Projection \citep{mcinnes2018umap}. These methods emphasize different recovered structures: Isomap is motivated by geodesic distance, Locally Linear Embedding and Local Tangent Space Alignment use local linear or tangent information, Hessian Eigenmaps uses local second-order structure, Laplacian Eigenmaps and Diffusion Maps recover spectral structure, and UMAP emphasizes neighborhood relations. Thus, methods such as UMAP are not intended to preserve geodesic distance in the same sense as Isomap, but rather to retain neighborhood or graph-based structure. Their theoretical guarantees are therefore method-dependent and concern different notions of recovery under different assumptions; see, for example, \citet{hein2007graph} for graph Laplacian convergence and \citet{meilua2024manifold} for a recent review. Accordingly, although both embedding and fitting are motivated by low-dimensional structure, embedding targets coordinates in a representation space whereas fitting targets an estimated manifold in the ambient space. Moreover, these embeddings usually represent a one-way transformation: operations performed in the lower-dimensional space cannot easily be translated back to the original high-dimensional ambient space.

\begin{figure}[htbp]
    \centering
    \resizebox{0.475\textwidth}{!}{
    \begin{tikzpicture}[x=0.75pt,y=0.75pt,yscale=-1,xscale=1]
        \clip (0, 0) rectangle (400, 150);
        \pgfmathsetseed{2024}
        
        \fill[gray!10] (0,110) to[bend right=20] (195,30) 
                       to[bend right=20] (390,110)
                       to[bend left=15] (195,115)
                       to[bend left=15] (0,110);
                       
        \draw [dashed] (0,110) to[bend right=20] (195,30) 
                       to[bend right=20] (390,110)
                       to[bend left=15] (195,115)
                       to[bend left=15] (0,110);
        \foreach \i in {1,...,25} {
          \pgfmathsetmacro{\x}{rand*150 + 195}
          \pgfmathsetmacro{\y}{rand*70 + 80}
          \fill (\x, \y) circle (2pt);
          }
        \pgfmathsetseed{2024}
        \foreach \i in {1,...,25} {
          \pgfmathsetmacro{\x}{rand*150/1.1 + 195}
          \pgfmathsetmacro{\y}{rand*70/1.5 + 80}
          \fill [red] (\x, \y) circle (2pt);
          }
    \end{tikzpicture}
    }
    \caption{Manifold Denoising: This illustration shows observed data points (black) distributed around a latent manifold (gray surface). Manifold denoising aims to adjust these points to the denoised positions (red points), bringing them closer to the latent manifold.}
    \label{fig:Illus-denoising}
\end{figure}
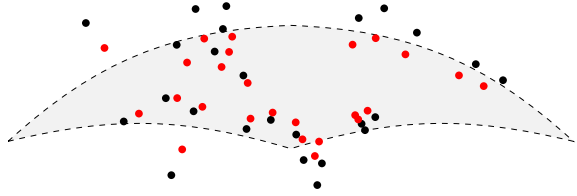

\noindent
\textit{Manifold denoising} targets the correction of noisy observations that are distributed around a low-dimensional manifold. A simple statistical setup is
\begin{equation*}
    Y_i = X_i + \varepsilon_i, \qquad X_i \in \mathcal{M},
\end{equation*}
where $X_i$ denotes an unobserved signal point on the latent manifold $\mathcal{M}$, $\varepsilon_i$ is a noise term, and only the noisy observation $Y_i$ is observed. The goal of denoising is to estimate corrected points $\tilde X_i$ from the observations $Y_i$ so that $\tilde X_i$ is closer to the latent manifold and better approximates $X_i$. In this sense, manifold denoising primarily targets pointwise recovery rather than direct estimation of the full manifold itself. There are two principal approaches to manifold denoising: feature-based and expectation-based methods.

Feature-based methods involve extracting features using techniques such as wavelet transformations \citep{deutsch2016manifold, yang2023manifold} or neural networks \citep{luo2020differentiable}. These methods eliminate non-informative features and employ inverse transformations to produce denoised data points. Expectation-based methods, on the other hand, achieve denoising by adjusting the local sample mean \citep{wang2010manifold} or fitting a local mean function \citep{sober2020manifold}. Although these approaches can effectively reduce noise, their theoretical guarantees are often method-specific or developed under restrictive assumptions. More importantly for the present review, they mainly target pointwise recovery and are less directly suited to reconstructing a smooth manifold object with explicit geometric regularity.

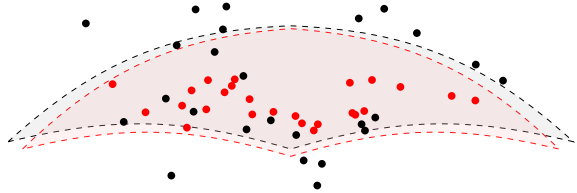
\begin{figure}[htbp]
    \centering
    \resizebox{0.475\textwidth}{!}{
    \begin{tikzpicture}[x=0.75pt,y=0.75pt,yscale=-1,xscale=1]
        \clip (0, 0) rectangle (400, 150);
        \pgfmathsetseed{2024}
        
        \fill[gray!10] (0,110) to[bend right=20] (195,30) 
                       to[bend right=20] (390,110)
                       to[bend left=15] (195,115)
                       to[bend left=15] (0,110);
                       
        \draw [dashed] (0,110) to[bend right=20] (195,30) 
                       to[bend right=20] (390,110)
                       to[bend left=15] (195,115)
                       to[bend left=15] (0,110);
                       
        \draw [dashed,red] (10,115) to[bend right=20] (195,32) 
                       to[bend right=20] (380,115)
                       to[bend left=15] (195,120)
                       to[bend left=15] (10,115);
       \fill[pink, opacity=0.2] (10,115) to[bend right=20] (195,32) 
                       to[bend right=20] (380,115)
                       to[bend left=15] (195,120)
                       to[bend left=15] (10,115);
        \foreach \i in {1,...,25} {
          \pgfmathsetmacro{\x}{rand*150 + 195}
          \pgfmathsetmacro{\y}{rand*70 + 80}
          \fill (\x, \y) circle (2pt);
          }

        \pgfmathsetseed{2024}
        \foreach \i in {1,...,25} {
          \pgfmathsetmacro{\x}{rand*150/1.15 + 195}
          \pgfmathsetmacro{\y}{rand*20 + 85}
          \fill [red] (\x, \y) circle (2pt);
          }
    \end{tikzpicture}
    }
    \caption{Manifold Fitting: This illustration shows observed data points (black) distributed around a latent manifold (gray surface). Manifold fitting seeks to construct a smooth manifold estimator (light red surface) that captures the essential properties of the latent manifold, enabling the projection of observations onto it (red points) for enhanced downstream analysis.}
    \label{fig:Illus-fitting}
\end{figure}

\vspace{8pt}
\noindent
\textit{Manifold fitting} is a crucial yet challenging aspect of learning the latent manifold. This method aims to estimate a smooth manifold $\widehat{\mathcal{M}}$ within the same ambient space that faithfully reflects the geometry and topology of an unknown low-dimensional manifold using data located on or near it. Unlike manifold embedding, which outputs coordinates in another space, or manifold denoising, which directly targets corrected points, manifold fitting first targets the manifold object itself. A fitted point $\widehat X_i$ is then obtained only after estimating $\widehat{\mathcal{M}}$, typically by projecting an observation onto the estimated manifold. The primary objective is therefore the recovery of $\widehat{\mathcal{M}}$, together with desirable geometric properties such as smoothness and accurate local structure.

One significant benefit of manifold fitting is its ability to disclose the shape of the concealed manifold through the estimator $\widehat{\mathcal{M}}$, after which one may project data samples onto the fitted manifold to obtain fitted points for downstream analysis. For instance, in biochemistry, manifold fitting is employed to reconstruct the three-dimensional structure of protein molecules from multiple cryo-electron microscopy (cryo-EM) images taken at various angles. Despite the high-dimensional noise in these images due to pixel scale, manifold fitting successfully recovers the underlying low-dimensional Lie group that corresponds to the molecules' orientations. Additionally, manifold fitting is useful in other areas, such as light detection and ranging \citep{kim2021nanophotonics} and wind direction detection \citep{dang2015wind}, where it aids in uncovering complex data patterns. Furthermore, manifold fitting can generate manifold-valued data with specific distributions, which is especially beneficial in developing generative machine learning models like Generative Adversarial Networks \citep{goodfellow2014generative}.

\section{Preliminary}
\subsection{Notations and mathematical concepts}
\label{Section:Notation}
Throughout this review, we utilize both upper- and lower-case $C$ to denote absolute constants. Upper-case $C$ typically represents constants greater than one, while $c$ denotes constants less than one. The values of these constants may change depending on the context.
In our notation, $x$ denotes a point on the latent manifold $\mathcal{M}$, $y$ represents a point associated with the observation, and $z$ indicates an arbitrary point of interest in the ambient space. The symbol $r$ is used to specify the radius in certain contexts. 
Mathematical entities related to sets are denoted using capitalized calligraphy letters, such as $\mathcal{M}$ for the manifold, $\mathcal{Y}$ for the observation set, and $\mathcal{B}_D(z, r)$ for a $D$-dimensional Euclidean ball centered at $z$ with radius $r$.

The distance between a point $a$ and a set $\mathcal{A}$ is given by $\nd(a,\mathcal{A}) = \min_{a^\prime \in \mathcal{A}} \|a-a^\prime\|_2$, where $\|\cdot\|_2$ is the Euclidean norm. To measure the discrepancy between two sets, we employ the \emph{Hausdorff distance}, which is frequently used in the assessment of estimator accuracy. This metric is particularly useful in evaluating the distance between the latent manifold $\mathcal{M}$ and its estimate $\widehat{\mathcal{M}}$. The Hausdorff distance is defined as follows:
\begin{definition}
    [Hausdorff distance] Let $\mathcal A$ and $\mathcal B$ be two non-empty subsets of $\bR^D$. Their Hausdorff distance $\dhaus(\mathcal A,\mathcal B)$ induced by Euclidean distance is defined as
    $$
    \dhaus(\mathcal A,\mathcal B) = \max \{\sup_{a\in \mathcal A} \inf_{b\in \mathcal B}\|a-b\|_2,~\sup_{b\in \mathcal B} \inf_{a\in \mathcal A}\|a-b\|_2\}.
    $$
\end{definition}
\begin{remark}
    For any $\mathcal A,~\mathcal B \subset \bR^D$,  $\dhaus(\mathcal A,\mathcal B)<\epsilon$ is equivalent to the fact that, for $\forall a\in \mathcal A$ and $\forall b\in \mathcal B$, $\nd(a,\mathcal B) <\epsilon$  and $\nd(b,\mathcal A) <\epsilon.$
\end{remark}
In the context of geometry, the Hausdorff distance provides a measure of the proximity between two manifolds. It is commonly acknowledged that a small Hausdorff distance implies a high level of alignment between the two manifolds, with controlled discrepancies.

We also require some essential geometrical concepts pertinent to the study of manifold fitting. A $d$-dimensional topological manifold is defined as a second-countable, Hausdorff topological space that is locally Euclidean of dimension $d$. Specifically, this implies that every point has a neighborhood homeomorphic to an open subset of $\mathbb{R}^d$. The tangent space at a point $x$ in a manifold $\mathcal{M}$, denoted $T_x\mathcal{M}$, is a $d$-dimensional affine space comprising all vectors tangent to $\mathcal{M}$ at $x$. A Riemannian metric $g$ on $\mathcal{M}$ is a smoothly varying collection of inner products on the tangent spaces, with each inner product $g_x: T_x\mathcal{M} \times T_x\mathcal{M} \to \mathbb{R}$ at $x \in \mathcal{M}$. Therefore, a Riemannian manifold is a pair $(\mathcal{M}, g)$, where $\mathcal{M}$ is a smooth manifold and $g$ is a Riemannian metric. In our context, we assume $\mathcal{M} \subset \mathbb{R}^D$ with $g$ induced by the Euclidean metric of $\mathbb{R}^D$, thereby simplifying $\mathcal{M}$ to a $d$-dimensional Riemannian manifold.

Projection matrices $\Pi_x^-$ and $\Pi_x^\perp$ project any vector $v \in \mathbb{R}^D$ onto the tangent space $T_x\mathcal{M}$ and its orthogonal complement, respectively. These matrices satisfy the relation $\Pi_x^\perp = I_D - \Pi_x^-$, where $I_D$ is the identity matrix in $\mathbb{R}^D$, and we denote estimators for these projections as $\widehat{\Pi}_z^\perp$ and $\widehat{\Pi}_z^-$. For an arbitrary point $z \not\in \mathcal{M}$, its projection onto the manifold is defined by $z^* = \arg\min_{x \in \mathcal{M}} \|x - z\|_2$.

The curvature of $\cM$ is described by the \emph{reach} of $\cM$. The concept of reach, as introduced by \citet{federer1959curvature}, is pivotal in assessing the regularity of manifolds embedded in Euclidean space and finds extensive applications in signal processing and machine learning. It can be defined as follows:
\begin{definition}
    [Reach] Let $\mathcal A$ be a closed subset of $\bR^D$. The reach of $\mathcal A$, denoted by $\reach(\mathcal A)$, is the largest number $\tau$ to have the following property: any point at a distance less than $\tau$ from $\mathcal A$ has a unique nearest point in $\mathcal A$.
\end{definition}
\begin{remark}
    The value of $\reach(\cM)$ can be interpreted as a second-order differential quantity if $\cM$ is treated as a function. Namely, let $\gamma$ be an arc-length parameterized geodesic of $\cM$; then, according to \citet{niyogi2008finding}, $\|\gamma^{\prime\prime}(t)\|_2\leq \reach(\cM)^{-1}$ for all $t$.
\end{remark}
Positive reach guarantees a tubular neighborhood around $\cM$ in which the nearest-point projection is uniquely defined and stable for points sufficiently close to the manifold. Consequently, it excludes sharp folds, self-near-intersections, and corner-type singularities. This makes reach especially useful in manifold fitting, where one often needs not only setwise closeness but also reliable projection and tangent-space approximation in the ambient space. In particular, positive reach is stronger than mere Hausdorff closeness: a set may be close to a smooth manifold in Hausdorff distance while still developing nonsmooth edges or vertices, and many piecewise linear approximations have zero reach at such locations. Reach is also an extrinsic geometric regularity quantity in Euclidean space, so it differs from intrinsic curvature notions such as Alexandrov, Ricci, or sectional curvature.

For example, the reach of a circle is its radius, and the reach of a linear subspace is infinite. Intuitively, a large reach implies that the manifold is locally close to the tangent space. This phenomenon can be explained by the following lemma given by \citet{federer1959curvature}:
\begin{lemma}[Federer's reach condition]
    \label{Lemma:ReachCond}
     Let $\cM$ be an embedded submanifold of $\bR^{D}$. Then,
    $$\reach(\cM)^{-1} = \sup \left\{\frac{2\textnormal{d} \mathit{(b,T_a\cM)}}{\|\mathit{a-b}\|_2^2} \mid \mathit{a,b}\in\cM,~\mathit{a\neq b}\right\}.$$
\end{lemma}

Calculating the average of a group of points on a manifold is not straightforward. To address this, we introduce the concept of the \emph{Fr\'echet mean}, which generalizes the idea of centroids to metric spaces, providing a representative point or central tendency for a cluster of points.
\begin{definition}
    [Fr\'echet Mean] Let $\{x_1, \dots, x_N\}$ be a collection of points on a manifold $\cM$. For any point $z$ on $\cM$, define the Fr\'echet function to be the sum of squared distances from $z$ to each $x_i$:
    \[
        \Psi(z) = \sum_{i=1}^N g^2(z, x_i).
    \]
    The set of points $\{\mu_F \in \cM : \Psi(\mu_F) = \arg\min_{z \in \cM} \Psi(z)\}$, where $\Psi(z)$ is minimized, is called the Fr\'echet mean set. If the minimizer is unique, $\mu_F$ is simply called the Fr\'echet mean of the set $\{x_1, \dots, x_N\}$ on $\cM$.
\end{definition}

\subsection{Model setting of manifold fitting}
The primary goal of manifold fitting is to construct a smooth manifold estimator from a set of noisy observations in the ambient space. This estimator aims to not only approximate the true manifold with a bounded geometric error but also to preserve its inherent geometric properties. To facilitate the development of a corresponding theoretical framework, we employ a widely recognized model setting, detailed in this subsection.

Throughout this review, we consider a random vector $Y \in \mathbb{R}^D$ modeled as
\begin{equation}
    \label{eq:Add_model}
    Y = X + \xi,
\end{equation}
where $X \in \mathbb{R}^D$ is an unobserved random vector whose law $\omega$ is supported on the unknown latent manifold $\mathcal{M}$, and $\xi \sim \phi_\sigma$ denotes the ambient-space observation noise, independent of $X$, characterized by a noise level $\sigma$. The distribution of $Y$ can be viewed as the convolution of $\omega$ and $\phi_\sigma$, with the density at a point $y$ given by
\begin{equation}
    \label{eq:def:nu}
    \nu(y) = \int_\cM \phi_\sigma(y-x)\, \mathrm{d}\omega(x).
\end{equation}
Equation~\eqref{eq:Add_model} should therefore be read not only as a geometric statement that data lie near a manifold, but also as a statistical generative model. The unknown latent manifold $\mathcal{M}$ carries a measure or distribution $\omega$ describing how the clean points are distributed along it, and the observed sample is informative only through the noisy law of $Y$ induced by $\omega$ and $\phi_\sigma$. In this sense, manifold fitting seeks recovery of an ambient-space manifold under an unknown latent-manifold-plus-noise generative model. The latent manifold is also typically unknown in manifold denoising, but denoising primarily targets cleaned points or local geometric summaries rather than a global smooth manifold estimator. This perspective also clarifies the taxonomy in Table~\ref{tab:manifold-paradigms}: manifold embedding and regularization need not rely on the explicit generative interpretation of Equation~\eqref{eq:Add_model}, whereas denoising and fitting are more directly tied to it, and manifold diffusion may lie between these viewpoints depending on formulation.

Assume $\mathcal{Y} = \{y_i\}_{i=1}^N \subset \mathbb{R}^D$ represents the collection of observed data points, each in the form of
\begin{equation}
    y_i = x_i + \xi_i, \quad \text{ for } i = 1,\cdots,N, 
\end{equation}
with $(y_i, x_i, \xi_i)$ being $N$ independent and identically distributed realizations of $(Y, X, \xi)$. Utilizing $\mathcal{Y}$, manifold fitting aims to construct an estimator $\widehat{\mathcal{M}}$ for $\mathcal{M}$ and provides theoretical justification for $\widehat{\mathcal{M}}$ under the following primary assumptions:
\begin{itemize}
    \item The latent manifold $\mathcal{M}$ is a compact, twice-differentiable $d$-dimensional submanifold, embedded within the ambient space $\mathbb{R}^D$. Its volume, with respect to the $d$-dimensional Hausdorff measure, is upper bounded by $V$, and its reach is lower bounded by a fixed constant $\tau$.
    \item The distribution $\omega$ is assumed to be supported on $\mathcal{M}$, and the density with respect to the $d$-dimensional Hausdorff measure is lower and upper bounded by non-zero constants.
    \item The intrinsic dimension $d$ of $\mathcal{M}$ and the noise level $\sigma$ are known parameters.
\end{itemize}

In many theoretical studies, the intrinsic dimension $d$ is treated as known. In practice, however, it often must be estimated from the observed data, and this task becomes substantially more difficult in the presence of noise. For users of manifold fitting methods, this issue is important because the estimated dimension directly influences neighborhood selection, local PCA or tangent-space estimation, bandwidth or scale choices, and therefore the stability of the final manifold estimator. Consequently, dimension estimation should often be regarded as a meaningful preliminary step in practical fitting pipelines rather than as a fixed input supplied in advance.

\section{Manifold fitting methods}
\subsection{Inspirations from traditional non-parametric methods}
The idea of fitting the latent manifold has significantly evolved since the 1980s, with one foundational approach being the Delaunay triangulation \citep{lee1980two}. In this technique, a mesh is created such that no sample points lie inside the circumcircle of any triangle in the triangulation. Early methods, as described by \citet{cheng2005manifold} and \citet{boissonnat2009manifold}, rely on dense samples free of noise, effectively creating an $(\epsilon,\delta)$-net of the hidden manifold. These approaches typically produce a piecewise linear manifold that is geometrically and topologically similar to the hidden manifold but lack smoothness. Moreover, the assumption of a noise-free and densely distributed dataset limits the widespread application of these algorithms.

This ambient-space objective differs from the classical manifold embedding literature, including Isomap, Locally Linear Embedding, Hessian Eigenmaps, Laplacian Eigenmaps, Diffusion Maps, Local Tangent Space Alignment, and UMAP, which instead construct low-dimensional coordinates adapted to different geometric, neighborhood, or spectral structures.

Additionally, a neighboring line of work seeks nonlinear or intrinsic extensions of PCA. In Euclidean space, \citet{donnell1994analysis} study smallest additive principal components. For a random vector $Y=(Y_1,\ldots,Y_D)$, they consider centered componentwise transformations $\phi_i(Y_i)$ and minimize $\operatorname{Var}\{\sum_i \phi_i(Y_i)\}$ subject to a normalization constraint on the individual variances. The resulting level set $\sum_i \phi_i(y_i)=0$ defines a co-dimension one additive manifold, which can be viewed as removing variation along a smallest nonlinear principal component. Around the same period, \citet{hastie1989principal} introduce principal curves, a nonlinear analogue of the first principal component: a smooth one-dimensional curve passing through the middle of the data cloud and summarizing its dominant mode of variation. Their construction starts from the linear PCA direction through the sample mean and iteratively updates the curve by conditional expectation until a self-consistent mean curve is obtained. These methods are important precursors because they move beyond linear subspaces, but they remain primarily PCA-type summaries of variation in Euclidean data rather than general ambient-space manifold estimators.

Related ideas also appear when the relevant manifold structure is already known. \citet{panaretos2014principal} propose principal flow, which constructs a smooth curve on a given Riemannian manifold by following the leading direction obtained from local tangent-space PCA, starting from the sample Fr\'echet mean; see Figure~4. In this sense, principal flow summarizes dominant variation on a known manifold rather than estimating an unknown manifold from noisy ambient observations. This viewpoint has led to several further developments, including principal boundary for manifold classification \citep{yao2020principal}, fixed boundary flow with prescribed endpoints \citep{yao2024random}, and principal submanifolds obtained by varying the initial tangent direction so that a family of principal-flow trajectories forms a higher-dimensional object \citep{yao2024principalsubmanifolds}. Other PCA-type generalizations on known nonlinear spaces include geodesic PCA and intrinsic shape analysis on Riemannian manifolds or quotient spaces \citep{huckemann2010intrinsic}, as well as principal nested spheres for spherical data \citep{principal_nested_spheres}. Moving from fixed nonlinear spaces toward data-adaptive nested hierarchies, principal nested submanifolds \citep{su2025principal} extend the nested-PCA idea to a sequence of smooth submanifolds embedded in the ambient space, thereby providing a PCA-like decomposition that is closer in spirit to manifold fitting. Compared with the main manifold-fitting methods reviewed below, however, these PCA-type methods primarily aim to summarize dominant variation through curves, flows, geodesic components, or nested submanifold structures, rather than to solve the general problem of recovering an unknown latent manifold under the additive noise model.

\begin{figure}
    \centering
    \subfloat{\includegraphics[width=0.475\textwidth]{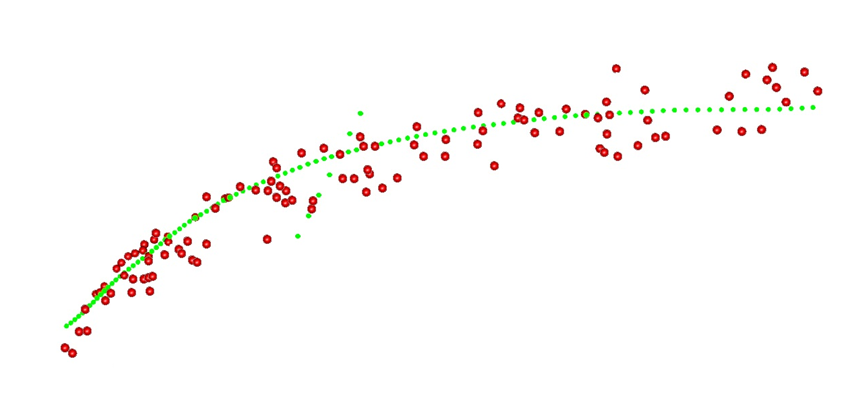}}
    \subfloat{\includegraphics[width=0.475\textwidth]{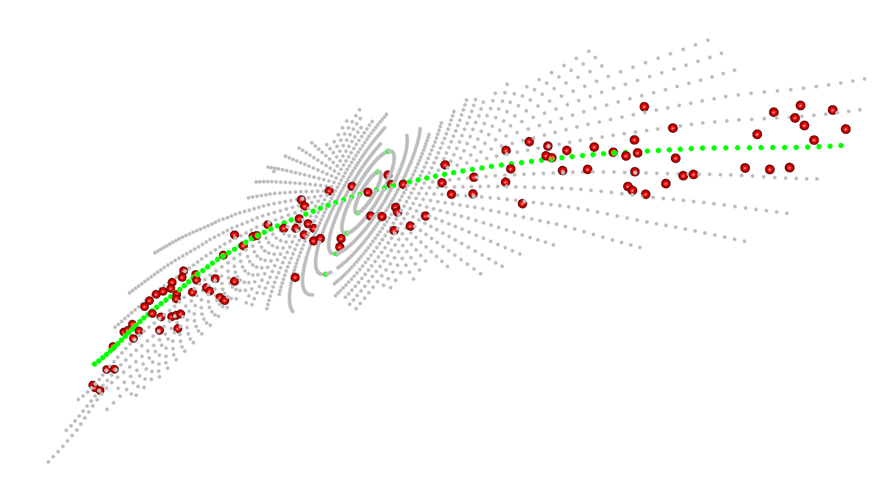}}
    \caption{Left panel: principal flow on a known manifold, starting from the sample Fr\'echet mean and evolving along the leading principal direction obtained from tangent-space PCA. Right panel: principal submanifold, obtained by varying the initial tangent direction so that a family of principal-flow trajectories forms a higher-dimensional object capturing multiple directions of variation.}
    \label{fig:enter-label}
\end{figure}

Moreover, some work focuses on local covariance structures and can perform on known manifolds. \citet{panaretos2014principal} propose a method named principal flow, fitting a 1-dimensional subspace on a known manifold. The curve $\gamma(t)$ starts from the sample Fr\'echet mean, and its derivative $\dot{\gamma}(t)$ aligns with the principal direction of the tangent space PCA along the curve. This method effectively represents the main variations of the samples on the manifold by leveraging PCA information from the tangent space of the known manifold. This technique has spurred further developments, such as principal boundary \citep{yao2020principal} for classifying samples on the manifold by utilizing smaller eigenvalue boundaries, and fixed boundary flow \citep{yao2024random} with designated starting and ending points on the manifold. Furthermore, by altering the starting direction of the principal flow within the tangent space, different trajectories can be obtained, and the collection of these trajectories forms what is known as the principal submanifold \citep{yao2024principalsubmanifolds}. By exploiting the local covariance structure of the samples, these methods offer the flexibility to modify the portrayal of sample variations. However, their reliance on predefined manifold structures and the iterative computation of tangent space PCA limits their broader applicability.

In recent years, the methodology for fitting the latent manifold has been refined to accommodate various types of noise and sample distributions, improving the smoothness of the resulting manifolds and broadening the scope of practical applications.

\subsection{Origins of manifold fitting}
A series of influential studies by \citet{genovese2012minimax, genovese2012manifold} explore manifold estimation through the lens of minimax risk under the Hausdorff distance, utilizing Le Cam's method. Their foundational paper \citep{genovese2012minimax} models noisy sample points as a combination of latent variables from a hidden manifold and additive noise, where the noise norm is bounded by a constant $\sigma > 0$ and is assumed perpendicular to the manifold. By constructing extreme cases, they establish that the optimal minimax estimation rate is lower bounded by $\mathcal{O}(N^{-2/(2+d)})$, and that the minimax risk is upper bounded by $\mathcal{O}\bigl(({N}/{\log N})^{-2/(2+d)}\bigl)$ through a sieve maximum likelihood estimator. This leads to the conclusion that the estimation rate is tightly bound, up to logarithmic factors, with an optimal rate of $\mathcal{O}(N^{-2/(2+d)})$. Remarkably, this rate depends only on the intrinsic dimension $d$, rather than the ambient dimension $D$. However, the noise assumption in their model is somewhat unrealistic, and the sieve maximum likelihood estimator lacks computational tractability.

Subsequent work by \citet{genovese2012manifold} extends the analysis to include noiseless, clutter noise, and additive noise models. In the additive model, noise is modeled more realistically as a general Gaussian distribution with isotropic standard deviation $\sigma$. They treat the sample distribution as a convolution of a manifold-valued distribution and a noise distribution, framing the fitting problem as one of deconvolution. They establish a lower bound for the optimal estimation rate at $\mathcal{O}\bigl(({\log N})^{-1}\bigl)$, and an upper bound that is a polynomial function of $({\log N})^{-1}$ using a standard deconvolution density estimator. However, the resulting output is not necessarily a manifold, and this approach relies on a known noise distribution, which is often impractical. Moreover, ensuring a small minimax risk necessitates an exponentially large sample size, which is not feasible in many applications.

Alongside these minimax analyses, earlier work in computational geometry and statistics had already clarified several ingredients that later became central to manifold fitting. In particular, \citet{chazal2008smooth} and \citet{Aizenbud_Sober} study reconstruction or estimation of manifolds from noisy observations under geometric regularity and neighborhood control, while related work such as \citet{aamari2019nonasymptotic} develops nonasymptotic estimation of local geometric quantities, especially tangent spaces. These contributions provide important background for later manifold fitting methods that explicitly target smooth manifold estimation in ambient space, even though they do not formulate exactly the same fitting problem in the sense considered by the later smooth ambient-space estimators reviewed here.

To circumvent the requirement of large sample sizes for consistent manifold estimation, \citet{genovese2014nonparametric} propose studying the ridge of the sample distribution as a proxy. They demonstrate that the Hausdorff distance between the ridge of the kernel density estimator (KDE) and the sample density's ridge is $\mathcal{O}_P\bigl(({N}/{\log N})^{-2/(8+D)}\bigl)$. Further, they show that the ridge of the sample density is within $\mathcal{O}(\sigma^2\log(1/\sigma))$ of the true manifold in Hausdorff distance. They subsequently implement the mean-shift algorithm \citep{ozertem2011locally} to estimate the KDE ridge. Similar approaches are taken in \citep{chen2015asymptotic, mohammed2017manifold}, where ridge estimation techniques are further refined with guaranteed convergence.

While these manifold fitting approaches effectively manage minimax risk, ensuring sufficient smoothness of the estimators poses a significant challenge. The applicability of KDE-based methods in high-dimensional settings remains questionable. Moreover, these kernel-based methods require a progressively decreasing kernel bandwidth as the sample size increases, which may lead to inadequate capture of local geometric features. This problem often results in suboptimal convergence rates that are influenced more by the ambient dimensionality $D$ than by the intrinsic dimensionality $d$. Furthermore, the manifolds generated by these methods may exhibit small, complicated twists that do not match the true local geometry of the latent manifold. To overcome these limitations, substantial recent research focuses on ensuring that the output manifold exhibits a lower bounded reach, thereby improving the smoothness and geometric fidelity of the estimators. This emphasis reflects the fact that manifold fitting typically requires a regular ambient-space geometry with stable local projection behavior, not merely a set that is close in Hausdorff distance.

\subsection{Insights from mathematical analysis} 
From the mathematical perspective, the task of defining a smooth manifold from discrete sample points has been a focal point of interest, largely spurred by advancements in the generalization of the Whitney extension theorem. Originating from \citet{whitney1934analytic}, this theorem asserts that any smooth function on a closed subset of a manifold can be extended across the entire manifold. Recent solutions to the Whitney extension problem by \citet{fefferman2006whitney, fefferman2005sharp} have not only refined our approaches to data interpolation but also led to the formulation of the geometric Whitney problem \citep{fefferman2020reconstruction,fefferman2021reconstruction}. This problem examines under what conditions a smooth $d$-dimensional submanifold $\widehat{\mathcal{M}} \subset \mathbb{R}^D$ can approximate a set $\mathcal{A}$, assessing the accuracy in terms of distance and smoothness. To tackle these challenges, various mathematical approaches have been proposed, pushing the boundaries of geometric analysis and manifold reconstruction.

\begin{figure}[htbp]
    \centering
    \resizebox{.475\textwidth}{!}{
        \begin{tikzpicture}
            \draw (3,-0.8040) node[below] {$\cM$} arc (60:120:6);
            \draw [dotted] (-3,0) --(3,0) node[above left] {$T_{z^*}\cM$};
            \draw [densely dotted] (0,0.5) circle (1.5);
            \fill (0,0.5) node[above]{$z$} circle (1pt);
            \draw [densely dotted] (0,-1) -- (0,1.5);
            \draw [dotted,->] (0,0.5) -- (1.299,1.25) node[left=25pt,below=5pt]{$r$};
            
            \fill [blue] (-1.041889, -0.09115348) node[below]{$x_i$} circle (1pt);
            \draw [blue,densely dotted] (-2.5, -0.3482578) -- (1,0.2688866);
            \draw [blue,->](0,0.5) -- (0.0696764,0.1048455);

            \fill [red] (0.695935336485, -0.04049717) node[below]{$x_j$} circle (1pt);
            \draw [red,densely dotted] (-0.7245861814802, 0.3016111038197) -- (0.695935336485, -0.04049717);
            \draw [red,->](0,0.5) -- (-0.1603345777298,0.1481173725313);
        \end{tikzpicture}
    }
    \caption{Illustration of the ridge-based manifold recovery of \citet{mohammed2017manifold}. The black curve denotes the latent manifold $\cM$, $z$ is an evaluation point near $\cM$, and the dotted circle marks its local neighborhood of radius $r$. The colored sample points $x_i$ and $x_j$ contribute local tangent approximations (colored dotted lines), and the colored arrows indicate the corresponding normal-distance terms used to build the smooth surrogate function $f(z)$. The fitted manifold is then recovered as the ridge set of this aggregated function.}
    \label{Fig:PreviousWork:Mohammed2017}
\end{figure}
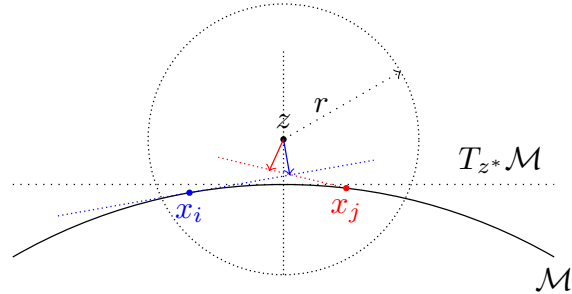

\subsubsection*{An early work without noise}
An important early contribution to manifold fitting in the absence of noise is presented by \citet{mohammed2017manifold}, which focuses exclusively on noiseless samples $\mathcal{X} = \{x_i \in \mathcal{M}\}_{i=1}^N$. In this work, the authors devise a method to reconstruct an estimator $\widehat{\mathcal{M}}$ using $\mathcal{X}$ by constructing a function $f(z)$ that approximates the squared distance from any point $z$ to $\mathcal{M}$. The ridge set of $f(z)$ is then utilized as an estimator of $\mathcal{M}$. The procedure is illustrated in Figure \ref{Fig:PreviousWork:Mohammed2017}. Figure \ref{Fig:PreviousWork:Mohammed2017} emphasizes that nearby samples contribute through local tangent-based distance surrogates, and the manifold is recovered as a ridge of their weighted aggregate.

For any point $z$ close to $\mathcal{M}$, its neighborhood index set within radius $r$ is defined as: 
$$I_z = \{i:\|x_i-z\|_2\leq r\}.$$
For each index $i \in I_z$, $\widehat{\Pi}_{x_i}^\perp$ is calculated using the smallest $D-d$ eigenvectors derived from local PCA. The squared distance from $z$ to the tangent space $T_{x_i}\mathcal{M}$ at $x_i$ is approximated by:
$$f_i(z)= \|\widehat\Pi_{x_i}^\perp(z-x_i)\|_2^2.$$
Then, $f(z)$ is designed as the weighted average of $f_i(z)$'s; that is, 
$$f(z) = \sum_{i\in I_z} \alpha_i(z) f_i(z),$$
where the weights $\alpha_i(z)$ are determined by bump functions:
$$\tilde{\alpha}_i(z)=\theta(\frac{\sqrt{f_i(z)}}{2r}),\quad \tilde{\alpha}(z) =  \sum_{i\in I_z}\tilde{\alpha}_i(z), \quad \alpha_i(z) = \frac{\tilde{\alpha}_i(z)}{\tilde{\alpha}(z)},$$
and $\theta(t)$ is a cutoff function such that $\theta(t)=1$ for $t\leq 1/4$ and $\theta(t) = 0$ for $t\geq 1$.

The estimator $\widehat{\cM}$ is given as the ridge set of $f(z)$; that is,
$$\widehat{\cM} = \{z\in\bR^D:~\nd(z,\cM)\leq cr,~\Pi_\textnormal{hi}(H_f(z))\partial f(z) = 0\},$$
where $H_f(z)$ is the Hessian matrix of $f$ at $z$, and $\Pi_\textnormal{hi}(A)$ projects matrix $A$ onto the span of the eigenvectors corresponding to its largest $D-d$ eigenvalues. This $\widehat{\mathcal{M}}$ is purported to have a reach bounded below by $cr$ and to be $\mathcal{O}(r^2)$-close to $\mathcal{M}$ in terms of Hausdorff distance.

The intuition behind this Hessian-based decomposition is simple. If $f$ behaves like an approximate squared distance to a manifold, then $f$ should curve sharply in directions normal to the manifold and much less in tangent directions. Accordingly, the large-eigenvalue eigenspace of $H_f(z)$ plays the role of a candidate normal space, and $\Pi_\textnormal{hi}$ projects onto that subspace; the complementary low-eigenvalue directions, which one may denote by $\Pi_\textnormal{lo}$, are therefore tangent-like directions. In this way, the Hessian separates ambient directions into normal-like and tangent-like components, which is the basic geometric mechanism behind the ridge viewpoint used here and in related putative-manifold formulations; see also \citet{mohammed2017manifold}. At a high level, this is also the intuition behind Fefferman-style putative-manifold constructions: second-order behavior identifies candidate normal directions, and the fitted manifold is then characterized by vanishing of the corresponding normal component of the gradient.

The condition $\Pi_\textnormal{hi}(H_f(z))\partial f(z)=0$ then means that the gradient has no component along the candidate normal directions, so $z$ lies on the central ridge-like set of $f$ that serves as the putative manifold.

Although this approach does not consider ambient space noise and depends significantly on accurately estimated projection directions $\widehat{\Pi}_{x_i}^\perp$, the method of approximating the distance function using projection matrices is innovative and sets a promising direction for future research. In particular, the method already moves beyond rough set approximation by constructing a smooth surrogate whose ridge defines the fitted object, and the lower bound on reach gives an explicit form of geometric regularity rather than mere Hausdorff closeness alone.

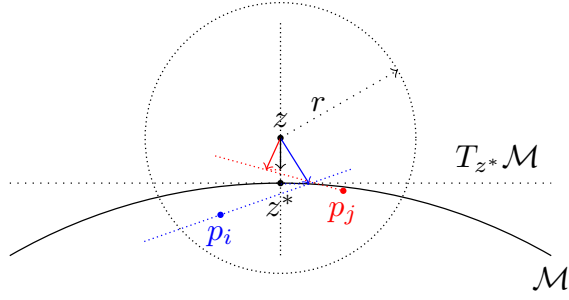
\begin{figure}[htbp]
    \centering
    \resizebox{.475\textwidth}{!}{
    \begin{tikzpicture}
        \draw (3,-0.8040) node[below] {$\cM$} arc (60:120:6);
        \draw [dotted] (-3,0) --(3,0) node[above left] {$T_{z^*}\cM$};
        \draw [densely dotted] (0,0.5) circle (1.5);
        \fill (0,0) node[below]{$z^*$} circle (1pt);
        \fill (0,0.5) node[above]{$z$} circle (1pt);
        \draw [dotted] (0,0) -- (0,0.5);
        \draw [dotted,->] (0,0.5) -- (1.299,1.25) node[left=25pt,below=5pt]{$r$};
        
        \fill [blue] (-0.6647, -0.3515) node[below]{$p_i$} circle (1pt);
        \draw [blue,densely dotted] (-1.510, -0.6469) -- (0.8,0.1603);
        \draw [blue,->](0,0.5) -- (0.3183,-0.0080);
        \draw [blue,densely dotted] (0,-0.0080)-- (0.3183,-0.0080);

        \fill [red] (0.695935336485, -0.0848142899562) node[below]{$p_j$} circle (1pt);
        \draw [red,densely dotted] (-0.7245861814802, 0.3016111038197) -- (0.695935336485, -0.0848142899562);
        \draw [red,->](0,0.5) -- (-0.1603345777298,0.1481173725313);
        \draw [red,densely dotted] (0,0.148117) -- (-0.1603345777298,0.1481173725313);

        \draw [->] (0,0.5) -- (0,0.15);
        \draw [densely dotted] (0,-0.8) -- (0,1.8);
    \end{tikzpicture}            
    }
    \caption{Illustration of the patching strategy proposed by \citet{fefferman2018fitting}. The black curve is the latent manifold $\cM$, $z^*$ is the nearest point on $\cM$ to the query point $z$, the horizontal dotted line is the tangent space $T_{z^*}\cM$, and the dotted circle marks a local neighborhood of radius $r$. The colored points $p_i$ and $p_j$ are nearby centers of sample-based local discs, the colored dotted lines indicate the orientations of these local geometric patches, and the arrows from $z$ represent their local bias contributions. The global estimator is obtained by patching these discs together through a smooth bias function, with reach used to control regularity.}
    \label{Fig:PreviousWork:Fefferman2018}
\end{figure}

\subsubsection*{An attempt with noise}
In a follow-up work, \citet{fefferman2018fitting} incorporate Gaussian noise from the ambient space into their analysis of manifold fitting. This study, building on the foundations laid by \citet{mohammed2017manifold}, aims to estimate the bias from an arbitrary point to the hidden manifold, with the collection of all zero-bias points serving as an estimator for $\mathcal{M}$.

To construct the bias function $f(z)$, the authors start with a sample set $\mathcal{Y}_0 = \{y_i\}_{i=1}^{N}$, ensuring the sample size meets specific criteria:
$$\frac{N}{\log(N)}>\frac{CV}{\omega_{\min}\beta_d(r^2/\tau)^d},\quad N\leq e^D,$$
where $V$ represents the volume of $\mathcal{M}$, $\beta_d$ is the volume of a Euclidean unit ball in $\mathbb{R}^d$, and $\omega_{\min}$ is the minimum value of $\omega$ on $\mathcal{M}$. Under these conditions, $\mathcal{Y}_0$ is $Cr^2/\tau$-close to $\mathcal{M}$ in Hausdorff distance with a probability of $1-N^{-C}$. From $\mathcal{Y}_0$, a subset $\mathcal{Y}_1 = \{p_i\}$ is selected greedily to form a minimal $cr/d$-net of $\mathcal{Y}_0$.

For each point $p_i$ in $\mathcal{Y}_1$, the authors define a $D$-dimensional ball $U_i = \mathcal{B}_D(p_i, r)$ and a $d$-dimensional ball $D_i = \mathcal{B}_d(p_i, r)$ centered at $p_i$. $D_i$ represents a disc cut from $U_i$, with its orientation determined by the \textit{FindDisc} algorithm developed by the authors. Ideally, $D_i$ should align parallel to $T_{p_i^*}\mathcal{M}$. The algorithm then estimates the basis of $D_i$ using sample points within $U_i$, which leads to an estimator $\widehat{\Pi}_{p_i}^\perp$ for the orthogonal projection.

For a point $z$ near $\mathcal{M}$, let $I_z = \{i : \|p_i - z\|_2 \leq r\}$, and define:
$$f_i(z) = \widehat{\Pi}_{p_i}^\perp(z-p_i),\quad \text{for } i \in I_z.$$
The bias function $f(z)$ is constructed as: 
\begin{align}\label{fy:fefferman18}
 f(z) = \sum_{i\in I_z} \alpha_i(z)(\widehat{\Pi}_z^\perp\widehat{\Pi}_{p_i}^\perp)(z-p_i),    
\end{align}
with $\widehat{\Pi}_z^\perp = \Pi_\textnormal{hi}\left(\sum_{i\in I_z} \alpha_i(z)\widehat{\Pi}_{p_i}^\perp\right)$, and the weights defined as 
$$\tilde{\alpha}_i(z)=\left(1 - \frac{\|z - p_i\|_2^2}{r^2}\right)^{d+2},\quad \tilde{\alpha}(z) =  \sum_{i\in I_z}\tilde{\alpha}_i(z), \quad \alpha_i(z) = \frac{\tilde{\alpha}_i(z)}{\tilde{\alpha}(z)},$$
for $z$ satisfying $\|z-p_i\|_2\leq r$ and $0$ otherwise. The resulting estimator $\widehat{\mathcal{M}}$ is defined as:
$$\widehat{\cM} = \{z\in\bR^D:~\nd(z,\cM)\leq cr,\quad f(z) = 0\}.$$

Setting $r = \mathcal{O}(\sqrt{\sigma})$, the authors demonstrate that $\widehat{\mathcal{M}}$ is $\mathcal{O}(\sigma)$-close to $\mathcal{M}$ and its reach is bounded below by $c\tau$ with probability $1-N^{-C}$. Notably, the algorithm for estimating disc orientation is not theoretically analyzed in the paper, and the accuracy of $f(z)$ is constrained by the successive projections $\widehat{\Pi}_z^\perp\widehat{\Pi}_{p_i}^\perp$ and the limited accuracy in estimating $\widehat{\Pi}_y^\perp$. Additionally, due to the constraints on the sample size $N$, the manifold estimation error retains a non-zero lower bound, limiting practical applications. Relative to earlier rougher approximations, the main contribution here is the explicit patching of local discs into a smooth ambient-space manifold together with projection stability encoded through the reach bound.

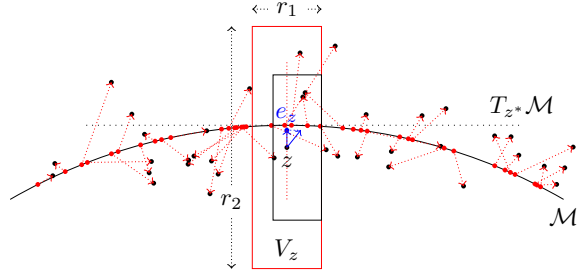
\begin{figure}[htbp]
    \centering
    \resizebox{.475\textwidth}{!}{
    \begin{tikzpicture}    
        \draw (0 ,0) node[below] {$\cM$} arc (60:120:8);
        \draw [dotted] (-6.9,1.0718) --(-4,1.0718) --(-0.6,1.0718) node[above] {$T_{z^*}\cM$};
        \draw (-4,-1) node[above] {$V_z$};
        \draw[red] (-4.5,-1) rectangle (-3.5,2.5);
        \draw[densely dotted, <->] (-4.5,2.75) --(-3.5,2.75) node[left = 6pt,fill=white] {$r_1$};
        \draw[densely dotted, <->] (-4.8,2.5) --(-4.8,-1) node[above = 21pt, fill=white] {$r_2$};
        \fill[red] (-2.8390,0.9871) circle (1pt);\fill (-2.9811,0.6126) circle (1pt);\draw[densely dotted, ->,red] (-2.8390,0.9871) --(-2.9811,0.6126);
        \fill[red] (-0.7018,0.3603) circle (1pt);\fill (0.1380,0.7562) circle (1pt);\draw[densely dotted, ->,red] (-0.7018,0.3603) --(0.1380,0.7562);
        \fill[red] (-4.9451,1.0158) circle (1pt);\fill (-5.1115,0.0816) circle (1pt);\draw[densely dotted, ->,red] (-4.9451,1.0158) --(-5.1115,0.0816);
        \fill[red] (-4.6281,1.0471) circle (1pt);\fill (-5.0900,0.5550) circle (1pt);\draw[densely dotted, ->,red] (-4.6281,1.0471) --(-5.0900,0.5550);
        \fill[red] (-1.5871,0.6993) circle (1pt);\fill (-1.4464,0.1926) circle (1pt);\draw[densely dotted, ->,red] (-1.5871,0.6993) --(-1.4464,0.1926);
        \fill[red] (-2.1841,0.8630) circle (1pt);\fill (-1.7007,1.3119) circle (1pt);\draw[densely dotted, ->,red] (-2.1841,0.8630) --(-1.7007,1.3119);
        \fill[red] (-6.5362,0.6591) circle (1pt);\fill (-5.9252,0.2342) circle (1pt);\draw[densely dotted, ->,red] (-6.5362,0.6591) --(-5.9252,0.2342);
        \fill[red] (-4.7549,1.0361) circle (1pt);\fill (-4.9937,0.3653) circle (1pt);\draw[densely dotted, ->,red] (-4.7549,1.0361) --(-4.9937,0.3653);
        \fill[red] (-0.3844,0.2081) circle (1pt);\fill (-0.1001,0.2172) circle (1pt);\draw[densely dotted, ->,red] (-0.3844,0.2081) --(-0.1001,0.2172);
        \fill[red] (-0.9954,0.4861) circle (1pt);\fill (-0.7597,0.9058) circle (1pt);\draw[densely dotted, ->,red] (-0.9954,0.4861) --(-0.7597,0.9058);
        \fill[red] (-3.9336,1.0715) circle (1pt);\fill (-3.8094,2.1171) circle (1pt);\draw[densely dotted, ->,red] (-3.9336,1.0715) --(-3.8094,2.1171);
        \fill[red] (-5.9049,0.8417) circle (1pt);\fill (-6.0577,0.9415) circle (1pt);\draw[densely dotted, ->,red] (-5.9049,0.8417) --(-6.0577,0.9415);
        \fill[red] (-4.6563,1.0448) circle (1pt);\fill (-4.1842,0.5993) circle (1pt);\draw[densely dotted, ->,red] (-4.6563,1.0448) --(-4.1842,0.5993);
        \fill[red] (-3.0402,1.0140) circle (1pt);\fill (-1.7753,0.7675) circle (1pt);\draw[densely dotted, ->,red] (-3.0402,1.0140) --(-1.7753,0.7675);
        \fill[red] (-0.7695,0.3905) circle (1pt);\fill (-0.9972,0.9182) circle (1pt);\draw[densely dotted, ->,red] (-0.7695,0.3905) --(-0.9972,0.9182);
        \fill[red] (-6.9658,0.5017) circle (1pt);\fill (-6.5370,1.6932) circle (1pt);\draw[densely dotted, ->,red] (-6.9658,0.5017) --(-6.5370,1.6932);
        \fill[red] (-3.7018,1.0662) circle (1pt);\fill (-3.7693,1.4801) circle (1pt);\draw[densely dotted, ->,red] (-3.7018,1.0662) --(-3.7693,1.4801);
        \fill[red] (-4.5874,1.0502) circle (1pt);\fill (-5.3435,0.6265) circle (1pt);\draw[densely dotted, ->,red] (-4.5874,1.0502) --(-5.3435,0.6265);
        \fill[red] (-4.2242,1.0687) circle (1pt);\fill (-3.4319,0.7174) circle (1pt);\draw[densely dotted, ->,red] (-4.2242,1.0687) --(-3.4319,0.7174);
        \fill[red] (-6.8839,0.5339) circle (1pt);\fill (-5.9580,0.5476) circle (1pt);\draw[densely dotted, ->,red] (-6.8839,0.5339) --(-5.9580,0.5476);
        \fill[red] (-2.9293,0.9998) circle (1pt);\fill (-2.8541,1.2264) circle (1pt);\draw[densely dotted, ->,red] (-2.9293,0.9998) --(-2.8541,1.2264);
        \fill[red] (-0.3336,0.1822) circle (1pt);\fill (-0.2795,0.4803) circle (1pt);\draw[densely dotted, ->,red] (-0.3336,0.1822) --(-0.2795,0.4803);
        \fill[red] (-0.8468,0.4241) circle (1pt);\fill (-0.6471,0.6389) circle (1pt);\draw[densely dotted, ->,red] (-0.8468,0.4241) --(-0.6471,0.6389);
        \fill[red] (-3.5170,1.0572) circle (1pt);\fill (-3.2643,0.6250) circle (1pt);\draw[densely dotted, ->,red] (-3.5170,1.0572) --(-3.2643,0.6250);
        \fill[red] (-2.3676,0.9035) circle (1pt);\fill (-2.4899,0.6930) circle (1pt);\draw[densely dotted, ->,red] (-2.3676,0.9035) --(-2.4899,0.6930);
        \fill[red] (-2.2436,0.8766) circle (1pt);\fill (-2.0460,1.2104) circle (1pt);\draw[densely dotted, ->,red] (-2.2436,0.8766) --(-2.0460,1.2104);
        \fill[red] (-4.0323,1.0717) circle (1pt);\fill (-3.3017,2.2170) circle (1pt);\draw[densely dotted, ->,red] (-4.0323,1.0717) --(-3.3017,2.2170);
        \fill[red] (-1.7272,0.7421) circle (1pt);\fill (-2.5288,0.4695) circle (1pt);\draw[densely dotted, ->,red] (-1.7272,0.7421) --(-2.5288,0.4695);
        \fill[red] (-5.6780,0.8938) circle (1pt);\fill (-5.1595,0.9918) circle (1pt);\draw[densely dotted, ->,red] (-5.6780,0.8938) --(-5.1595,0.9918);
        \fill[red] (-4.8404,1.0275) circle (1pt);\fill (-4.5391,1.6005) circle (1pt);\draw[densely dotted, ->,red] (-4.8404,1.0275) --(-4.5391,1.6005);
        \fill[red] (-7.5996,0.2162) circle (1pt);\fill (-7.3842,0.3415) circle (1pt);\draw[densely dotted, ->,red] (-7.5996,0.2162) --(-7.3842,0.3415);
        \fill[red] (-0.4140,0.2231) circle (1pt);\fill (0.0170,0.4563) circle (1pt);\draw[densely dotted, ->,red] (-0.4140,0.2231) --(0.0170,0.4563);
        \fill[red] (-7.2041,0.4021) circle (1pt);\fill (-7.3544,0.5188) circle (1pt);\draw[densely dotted, ->,red] (-7.2041,0.4021) --(-7.3544,0.5188);
        \fill[red] (-6.1143,0.7874) circle (1pt);\fill (-5.9907,0.3842) circle (1pt);\draw[densely dotted, ->,red] (-6.1143,0.7874) --(-5.9907,0.3842);
        \fill[red] (-4.7259,1.0388) circle (1pt);\fill (-5.4281,0.5091) circle (1pt);\draw[densely dotted, ->,red] (-4.7259,1.0388) --(-5.4281,0.5091);
        \fill[red] (-6.4395,0.6908) circle (1pt);\fill (-6.2409,1.2324) circle (1pt);\draw[densely dotted, ->,red] (-6.4395,0.6908) --(-6.2409,1.2324);
        \fill[red] (-3.1933,1.0310) circle (1pt);\fill (-3.7367,1.5424) circle (1pt);\draw[densely dotted, ->,red] (-3.1933,1.0310) --(-3.7367,1.5424);
        \fill[red] (-5.8119,0.8639) circle (1pt);\fill (-5.4203,0.5675) circle (1pt);\draw[densely dotted, ->,red] (-5.8119,0.8639) --(-5.4203,0.5675);

        \fill (-4,0.75) node[below]{$z$} circle (1pt);
        \fill [blue] (-4,1) node[above]{$e_z$} circle (1pt);
        \draw [blue,densely dotted] (-4,1) -- (-3.8,1);
        \draw [blue,->] (-4,0.75) -- (-4,1);       
        \draw [blue,->] (-4,0.75) -- (-3.8,1);
        \draw [red,densely dotted] (-4,0) -- (-4,2);     
        \draw (-4.2,-0.3) rectangle (-3.5, 1.8);
    \end{tikzpicture}           
    }
    \caption{Illustration of the noisy refinement strategy proposed by \citet{fefferman2021fitting}. The black curve denotes $\cM$, the dotted horizontal line indicates the local tangent direction, and the red elongated rectangle $V_z$ represents an anisotropic hyper-cylinder-type neighborhood with widths $r_1$ and $r_2$. Red points denote noisy observations, dotted red arrows indicate their local refinement within these regions, and the blue point $e_z$ is the resulting averaged or refined point associated with $z$. In contrast to Figure \ref{Fig:PreviousWork:Fefferman2018}, this step inserts a local denoising refinement before the final manifold patches are reconstructed.}
    \label{Fig:PreviousWork:Fefferman2021}
\end{figure}

\subsubsection*{An attempt to combine fitting with denoising}
Expanding on previous results, the error in fitting $\cM$ is typically upper bounded by two components: the distance from the sample to $\mathcal{M}$, which can generally be considered $\mathcal{O}(\sigma)$ as Gaussian noise tends to dissipate within several standard deviations, and the distance from $\widehat{\mathcal{M}}$ to the sample, which currently aligns tightly with the first component. Since the sampling bias $\dhaus(\mathcal{Y}, \mathcal{M}) = \mathcal{O}(\sigma)$ inhibits closer approximation to $\mathcal{M}$, denoising becomes essential to refine $\widehat{\mathcal{M}}$.

Building on \citet{fefferman2018fitting}, subsequent work by \citet{fefferman2021fitting} offers enhanced results through the refinement of points and the construction of a mesh grid on each disc $D_i$. As depicted in Figure \ref{Fig:PreviousWork:Fefferman2021}, each hyper-cylinder in the mesh extends significantly longer in the direction perpendicular to the manifold than parallel. Within these hyper-cylinders, a subset of $\mathcal{Y}_0$ is selected through a complex design, and their average is denoted by $e_y$. The aggregate of such $e_y$ across all hyper-cylinders forms $\mathcal{Y}_1$, which demonstrates proximity to $\mathcal{M}$ with a distance of $Cd\sigma^2/\tau$.

These refined points in $\mathcal{Y}_1$ are then used as input for the subsampling algorithm from \citet{fefferman2018fitting} to construct a new series of discs $\{D_i^\prime\}$. Utilizing both the refined points and discs, the same function $f(z)$ yields an $\widehat{\mathcal{M}}$ that is $\mathcal{O}(\sigma^2)$-close to $\mathcal{M}$ and maintains a reach of at least $c\tau$ with a probability of $1-N^{-C}$. Thus, the denoising step is used not merely to improve pointwise accuracy, but to support a smoother and more geometrically stable fitted manifold with the same regularity target as in the patching construction.

To date, the results presented by \citet{fefferman2021fitting} represent a state-of-the-art error bound for manifold fitting. However, there are several challenges in implementing the described method: The refinement step for $e_z$ requires sampling directly from the latent manifold, which violates the initial assumption of noisy data. In addition, the procedures for point refinement and disk orientation determination are only briefly outlined and may not be readily applicable to real-world data sets. Furthermore, similar to \citep{fefferman2018fitting}, the requirement for an upper bound on the sample size limits the practical application and asymptotic behavior of this algorithm.

\subsection{More practicable statistical methods}
Recently, statisticians have renewed their attention to improving the practicality of manifold fitting methods. They aim to eliminate the constraints imposed by the upper bound on the sample size while maintaining the smoothness of the manifold estimator. In addition, there is a growing interest in integrating manifold fitting techniques into traditional data analysis procedures. This integration is intended to provide more accurate and widely applicable nonlinear dimensionality reduction, thereby extending the utility of manifold fitting to various data-intensive domains.

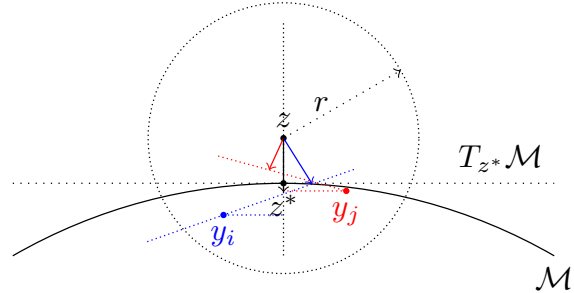
\begin{figure}[htbp]
    \centering
    \resizebox{.475\textwidth}{!}{
        \begin{tikzpicture}
            \draw (3,-0.8040) node[below] {$\cM$} arc (60:120:6);
            \draw [dotted] (-3,0) --(3,0) node[above left] {$T_{z^*}\cM$};
            \draw [densely dotted] (0,0.5) circle (1.5);
            \fill (0,0) node[below]{$z^*$} circle (1pt);
            \fill (0,0.5) node[above]{$z$} circle (1pt);
            \draw [dotted] (0,0) -- (0,0.5);
            \draw [dotted,->] (0,0.5) -- (1.299,1.25) node[left=25pt,below=5pt]{$r$};
            
            \fill [blue] (-0.6647, -0.3515) node[below]{$y_i$} circle (1pt);
            \draw [blue,densely dotted] (-1.510, -0.6469) -- (0.8,0.1603);
            \draw [blue,->](0,0.5) -- (0.3183,-0.0080);
            \draw [blue,densely dotted] (-0.6647, -0.3515) -- (0, -0.3515);
            \draw [red,densely dotted] (0.69593, -0.0848) -- (0, -0.0848);
    
            \fill [red] (0.695935336485, -0.0848142899562) node[below]{$y_j$} circle (1pt);
            \draw [red,densely dotted] (-0.7245861814802, 0.3016111038197) -- (0.695935336485, -0.0848142899562);
            \draw [red,->](0,0.5) -- (-0.1603345777298,0.1481173725313);
            \draw [->] (0,0.5) -- (0,-0.1);
            \draw [densely dotted] (0,-0.8) -- (0,1.8);
        \end{tikzpicture}                   
    }
    \caption{Illustration of the local bias construction proposed by \citet{yao2019manifold}. The point $z$ is a query point in the ambient space, $z^*$ is its nearest point on the latent manifold $\cM$, the horizontal dotted line is the tangent space $T_{z^*}\cM$, the vertical dotted line indicates the normal direction through $z$, and the dotted circle marks the local neighborhood of radius $r$. The colored observations $y_i$ and $y_j$ are nearby noisy samples; the colored dotted lines represent their local tangent estimates, while the colored arrows from $z$ illustrate the sample-based contributions used to estimate the normal projection and define the local bias function $f(z)$ that characterizes the fitted manifold.}
    \label{Fig:PreviousWork:Yao2019}   
\end{figure}

\subsubsection*{A more stable estimation for noisy data}
To address limitations related to sample size and tangent space estimation encountered in the work of \citet{fefferman2018fitting}, \citet{yao2019manifold} introduce an improved method that simplifies the estimation process. This method foregoes continuous projections, offering a more effective estimation of ${\Pi}_{z^*}^\perp$. The authors argue that accurately fitting the manifold is sufficient to estimate both the projection direction and the local mean effectively, as the manifold can be approximated as a linear subspace locally, with the local sample mean serving as a reliable reference point for the hidden manifold.

They start with a sample set $\mathcal{Y} = \{y_i\}_{i=1}^N$. For each $y_i$, $\widehat{\Pi}_{y_i}^\perp$ is derived using local PCA within a radius $r = \cO(\sqrt{\sigma})$, which has been demonstrated to produce satisfactory estimation errors. For any arbitrary point $z$, with $I_z = \{i : \|y_i - z\|_2 \leq r\}$, the bias function is constructed as:
\begin{align}\label{fy:yao2019}
f(z) = \widehat{\Pi}_z^\perp\left(z - \sum_{i\in I_z}\alpha_i(z)y_i\right),    
\end{align}
where $\widehat{\Pi}_z^\perp = \Pi_\textnormal{hi}\left(\sum_{i\in I_z}\alpha_i(z)\widehat{\Pi}_{y_i}^\perp\right)$. The weights are defined as:
$$\tilde{\alpha}_i(z) = \left(1 - \frac{\|z - y_i\|_2^2}{r^2}\right)^\beta,\quad \tilde{\alpha}(z) = \sum_{i\in I_z}\tilde{\alpha}_i(z), \quad \alpha_i(z) = \frac{\tilde{\alpha}_i(z)}{\tilde{\alpha}(z)},$$
for $z$ satisfying $\|z - y_i\|_2 \leq r$ and $0$ otherwise, with $\beta \geq 2$, a fixed integer that ensures second-order derivability of $f(z)$. This bias function defines the output manifold as:
$$\widehat{\cM} = \{z\in\bR^D:~\nd(z,\cM)\leq cr,\quad f(z) = 0\},$$
which is demonstrated to be $\mathcal{O}(\sigma)$-close to $\mathcal{M}$ in terms of Hausdorff distance and maintains a reach of at least $c\tau$ with a probability of $1 - c\exp(-Cr^{d+2}N)$.

Although the theoretical error bound remains consistent with that proved by \citet{fefferman2018fitting}, the method proposed by \citet{yao2019manifold} significantly simplifies the computational process and shows favorable numerical performance in many scenarios. Its main strength is that the fitted manifold is still characterized through a projection-based local bias function, so the method continues to target a smooth geometric object with stable local normal directions rather than only a denoised collection of points.

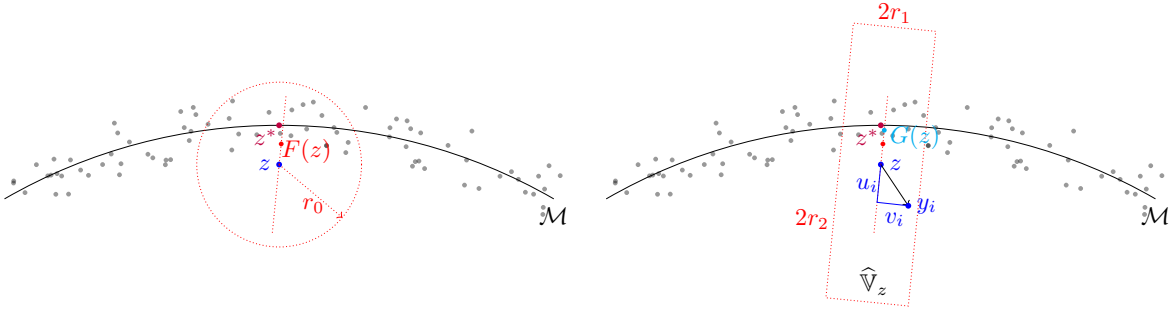
\begin{figure}[htbp]
        \centering
        \resizebox{0.475\textwidth}{!}{
        \begin{tikzpicture}
            \pgfmathsetmacro{\Centerx}{-8*cos(60)}
            \pgfmathsetmacro{\Centery}{-8*sin(60)}
            \clip (-8.2, -2) rectangle (0.2, 3);
            \draw (0 ,0) node[below] {$\mathcal{M}$} arc (60:120:8);
            \pgfmathsetseed{2025}
            \foreach \x in {1,...,80}{
                \pgfmathsetmacro{\Angle}{rand*(30)+90}
                \pgfmathsetmacro{\NoisyRadius}{8+rand*.4}
                \fill[shift={(\Centerx,\Centery)}, opacity=0.4] (\Angle:\NoisyRadius) circle (1pt);
            }
            \fill [purple] (-4,1.0718) circle (1.2pt);
            \node [purple] at (-4.2,0.9) {$z^*$};
            
            \fill [blue] (-4.0, 0.5) node[left]{$z$} circle (1.2pt);
            \draw [densely dotted,red] (-4.0,0.5) circle (1.2);
            \draw [red,densely dotted] (-4.0, 0.5) -- ++(0.1,1);
            \draw [red,densely dotted] (-4.0, 0.5) -- ++(-0.1,-1);
            \draw [densely dotted,->,red] (-4.0,0.5) -- ++(320:1.2);
            
            \fill [red, shift={(0.03,0.3)}] (-4.0, 0.5) circle (1pt);
            \node [red] at (-3.6, 0.7) {$F(z)$};
            \node [red] at (-3.5,-0.1) {$r_0$};
            
        \end{tikzpicture}
        }
        \resizebox{0.475\textwidth}{!}{
        \begin{tikzpicture}
            \clip (-8.2, -2) rectangle (0.2, 3);
            \draw (0 ,0) node[below] {$\mathcal{M}$} arc (60:120:8);
            \pgfmathsetmacro{\Centerx}{-8*cos(60)}
            \pgfmathsetmacro{\Centery}{-8*sin(60)}
            \pgfmathsetseed{2025}
            \foreach \x in {1,...,80}{
                \pgfmathsetmacro{\Angle}{rand*(30)+90}
                \pgfmathsetmacro{\NoisyRadius}{8+rand*.4}
                \fill[shift={(\Centerx,\Centery)}, opacity=0.4] (\Angle:\NoisyRadius) circle (1pt);
            }

            \fill [purple] (-4,1.0718) circle (1.2pt);
            \node [purple] at (-4.2,0.9) {$z^*$};
            
            \fill [blue] (-4.0, 0.5) node[right]{$z$} circle (1.2pt);
            \draw [red,densely dotted] (-4.0, 0.5) -- ++(0.1,1);
            \draw [red,densely dotted] (-4.0, 0.5) -- ++(-0.1,-1);

            \draw[red,densely dotted] (-4.6, 0.56) -- ++(0.2,2) -- ++(1.2,-0.12) -- ++(-0.4,-4) -- ++(-1.2,0.12) -- ++(0.2,2);
            
            \fill [red, shift={(0.03,0.3)}] (-4.0, 0.5) circle (1pt);
            \fill [cyan, shift={(0.05,0.5)}] (-4.0, 0.5) circle (1pt);

            \node [cyan] at (-3.5, 0.9) {$G(z)$};
            
            \node at (-4.1,-1.2) {$\widehat{\bV}_z$};
            
            \node[red] at (-3.8,2.7) {$2r_1$};
            \node[red] at (-5.0,-0.3) {$2r_2$};

            \draw [->] (-4.0, 0.5) -- (-3.6,-0.1);
            \fill [blue] (-3.6,-0.1) node[right]{$y_i$} circle (1.2pt);
            \draw [blue] (-3.6,-0.1) -- (-4.055,-0.05) -- (-4.0, 0.5);
            \node[blue] at (-3.8,-0.3) {$v_i$};
            \node[blue] at (-4.2, 0.2) {$u_i$};
        \end{tikzpicture}
        }
        \caption{Illustration of the denoising construction proposed by \citet{yao2023manifold}. Left panel: $z$ is the current point, $z^*$ is the latent target point on $\mathcal{M}$, the dotted circle of radius $r_0$ is the local neighborhood used to compute the weighted average $F(z)$ (red), and the direction of $F(z)-z$ estimates the direction from $z$ toward $z^*$. Right panel: the estimated direction from $F(z)-z$ is used to define the elongated anisotropic region $\widehat{\bV}_z$ with widths $2r_1$ and $2r_2$; a nearby observation $y_i$ is decomposed into the directional component $u_i$ and the orthogonal component $v_i$, and the resulting weighted contraction produces the denoised point $G(z)$ (cyan).}
        \label{Fig:Interpret_uv}
\end{figure}

\subsubsection*{A more flexible denoising module}
Building on the hyper-cylinder region concept introduced by \citet{fefferman2021fitting}, \citet{yao2023manifold} propose a specially designed method to enhance denoising, making it more practical and efficient. The method relies on an explicit geometric operator, {known as the {\it Yao--Yau estimator} (Theorem 4.5, \citet{yao2023manifold}). While it follows the general framework of \citep{yao2019manifold}, it simplifies the process by focusing on estimating a main direction rather than the entire projection matrix.} Compared with \citet{yao2019manifold}, the key geometric simplification is that the method estimates only the dominant direction from $z$ toward $z^*$ and then uses it to build an anisotropic denoising region. Specifically, the projection matrix onto $z^* - z$ is defined as:
$$U = {(z^* - z)(z^* - z)^\top}/{\|z^* - z\|_2^2}.$$
A narrow region can be constructed based on it as
\begin{equation}
    \label{eq:V_z:3}
    {\bV}_{z} = \cB_{D-1}(z,r_1) \times \cB_1(z,r_2),
\end{equation}
where the second term is an interval in the direction of $z^* - z$ with $r_2 = C\sigma\sqrt{\log(1/\sigma)}$, and the first term is in the span of the orthogonal complement of $z^* - z$ in $\bR^D$ with $r_1 = c\sigma$. The conditional expectation of $Y$ within ${\bV}_{z}$ serves as a denoised version of $z$.

To construct smooth estimators for these denoised points, the authors first use a local weighted average at point $z$ as the reference of $z^*$:
\begin{equation}
    \label{eq:def:F(z)}
    F(z) = \sum_{i}\alpha_i(z) y_i,
\end{equation}
with the weights being defined as
\begin{equation}
    \label{eq:def:weight_F}
\tilde{\alpha}_i(z)=
\left\{\begin{array}{cc}
\left(1 - \frac{\|z - y_i\|_2^2}{r_0^2}\right)^{k}, & \|z - y_i\|_2\leq r_0;\\
0, &  \textnormal{otherwise};\\ 
\end{array}\right.
\quad \tilde{\alpha}(z) =  \sum_{i\in I_z}\tilde{\alpha}_i(z),
\quad \alpha_i(z) = \frac{\tilde{\alpha}_i(z)}{\tilde{\alpha}(z)},
\end{equation}
with $k>2$ being a fixed integer to ensure $F(z)$ is twice-differentiable. The direction of $F(z) - z$ is shown to approximate the vector from $z$ to $z^*$ effectively. Consequently, a smooth estimator for $U$ is defined as:
$$\widehat{U} = \frac{(F(z)-z)(F(z)-z)^\top}{\|F(z)-z\|_2^2}.$$
For a data point $y_i$, the decomposition into orthogonal components is:
\begin{equation}
    \label{eq:def_u_i_v_i}
    u_i = \widehat{U}(y_i-z),\quad v_i = y_i - z - u_i.
\end{equation}
The contracted point of $z$ is given by:
\begin{equation}
    \label{eq:def:G(z)}
    G(z) = \sum_{i}\beta_i(z) y_i,  
\end{equation}
with the weights given by
\begin{equation}\label{eq:def:weight_G}
    \begin{aligned}
        w_u(u_i) &= \left\{
        \begin{array}{cl}
            1, & \|u_i\|_2\leq \frac{r_2}{2}; \\
            \left(1 - (\frac{2\|u_i\|_2-r_2}{r_2})^2\right)^k, & \|u_i\|_2\in (\frac{r_2}{2},r_2);\\
            0, & \textnormal{otherwise},
        \end{array}\right.\\
        w_v(v_i) &= \left\{
        \begin{array}{cl}
            \left(1 - \frac{\|v_i\|_2^2}{r_1^2}\right)^k, & \|v_i\|_2\leq r_1; \\
            0, & \textnormal{otherwise} ,
        \end{array}\right.\\
        \beta_i(z) = w_u(u_i)&w_v(v_i),
        \quad \tilde{\beta}(z) =  \sum\tilde{\beta_i}(z),
        \quad \beta_i(z) = \frac{\tilde{\beta_i}(z)}{\tilde{\beta}(z)},
    \end{aligned}
\end{equation}
where $k \geq 2$ ensures that $G(z)$ is also a $C^2$-continuous map from $\mathbb{R}^D$ to $\mathbb{R}^D$. The accuracy of $G(z)$ is shown to be upper bounded as $\|G(z) - z^*\|_2 \leq C\sigma^2{\log(1/\sigma)}$ with high probability.

The functions $F$ and $G$ significantly push the noisy points towards the latent manifold, which is the target of denoising. Moreover, if there exists a $d$-dimensional preliminary estimation $\widetilde{\cM}$ approximately $\cO(\sigma)$ close to $\cM$, $G(\widetilde{\cM})$ forms a $d$-dimensional manifold with an approximate error of $\cO(\sigma^2\log(1/\sigma))$ and a reach no less than $c\sigma\reach(\widetilde{\cM})$ with high probability.

This approach effectively exploits the geometric information inherent in the underlying manifolds by strategically selecting neighborhood radii and creating both spherical and rectangular neighborhoods. Building on the foundation established by \citet{yao2019manifold}, it also avoids the computational challenges associated with performing local PCA, thereby broadening the practical applicability of the method. This enhancement not only simplifies the computational process, but also broadens the circumstances where the method can be used effectively. Compared with the earlier patching-based methods, the emphasis here is more indirect: geometric regularity is promoted through anisotropic denoising and the smooth maps $F$ and $G$, while control of reach is inherited through the improvement of a preliminary manifold estimator rather than through an explicit curvature construction from the outset.

\begin{figure}[htbp]
    \centering
    \includegraphics[width=0.95\linewidth]{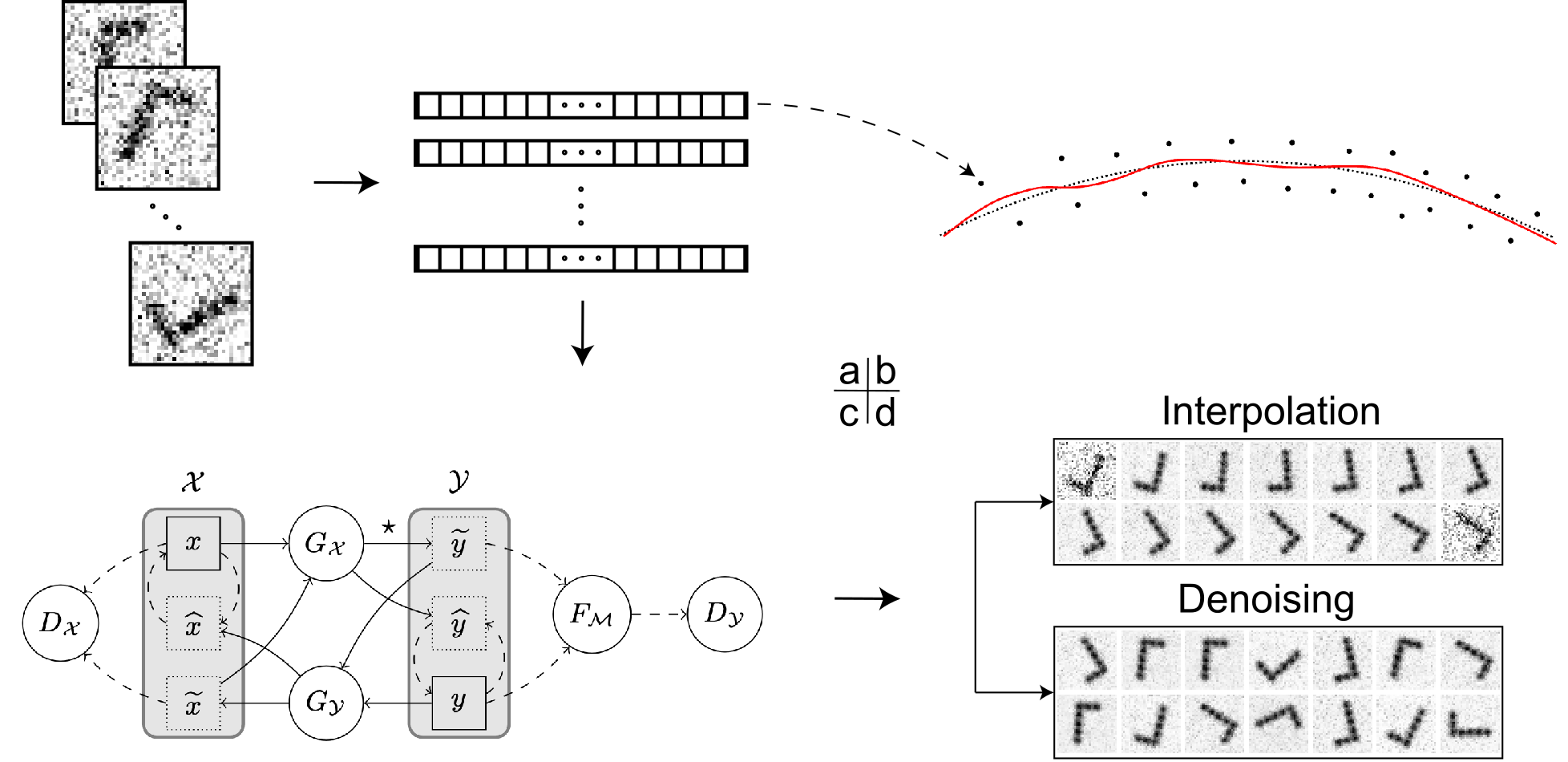}
    \caption{Illustration of fitting the latent manifold using the CycleGAN framework \citep{Zhu_2017_ICCV}. In the real world, data, such as the images shown in panel (a), are often treated as high-dimensional vectors. These vectors typically reside around a low-dimensional latent manifold, depicted by the black dotted curve in panel (b). The CycleGAN framework, detailed in panel (c), effectively learns to estimate this latent manifold (illustrated as the red curve in panel (b)). This construction supports nonlinear interpolation and projection-based denoising in the high-dimensional ambient space (panel (d)), with the aim of producing cleaner generated samples and smoother traversal along the learned latent structure.}
    \label{fig:mfcgan}
\end{figure}

\subsection{Scientific applications}
In addition to theoretical research in manifold fitting, statisticians have integrated manifold fitting concepts into broader data science applications, such as generative neural networks and bioinformatics. This integration enhances the information processing capabilities of the original pipelines while reducing the computational demands traditionally associated with manifold fitting methods.

\subsubsection*{Combination with generative neural networks}
Figure \ref{fig:mfcgan} illustrates the overview of the method proposed by \citet{MFCGAN}, which innovates by integrating manifold fitting with generative adversarial networks (GANs). A GAN is a deep learning architecture that involves two competing neural networks. One network, the generator, creates new data instances from the input features, attempting to mimic the real data as closely as possible. The other network, the discriminator, evaluates whether the output from the generator is authentic, meaning it could plausibly be part of the original data set. The process continues until the discriminator cannot reliably distinguish between real and generated data. This setup allows for the generation of increasingly realistic data, such as new images or audio compositions. Cycle Generative Adversarial Networks (CycleGAN) \citep{Zhu_2017_ICCV} extend this approach by simultaneously training two pairs of generators and discriminators to learn mappings between two different spaces. This capability makes it particularly useful for tasks such as photo conversion between different styles.

The approach proposed by \citet{MFCGAN} uses the CycleGAN framework of \citet{Zhu_2017_ICCV}, but introduces innovations in combining it with the concept of manifold fitting. Specifically, they use the low-dimensional feature space to control manifold dimensionality and introduce an additional noise component in the high-dimensional space. This modification helps to reduce the risk of overfitting. The generated data are then refined using a manifold fitting method proposed by \citet{yao2023manifold}, which acts as a denoising and projection step for generated samples and supports nonlinear interpolation along the learned manifold.

The integration of manifold fitting with CycleGAN provides two benefits. First, it allows the manifold dimensionality to be controlled using neural network architectures rather than traditional PCA, simplifying the computational process and enabling more efficient parallel computations. Second, it addresses the challenge of high noise levels in data generated by low signal-to-noise neural networks. Within this framework, manifold fitting is used to project generated samples back toward the learned manifold, thereby supporting noise reduction and smoother interpolation between latent representations. The reported experiments therefore point more specifically to favorable empirical performance in denoising noisy generated images and in nonlinear interpolation tasks, rather than to an undifferentiated gain in image quality.

\begin{figure}[htbp]
    \centering
    \includegraphics[width=0.95\linewidth]{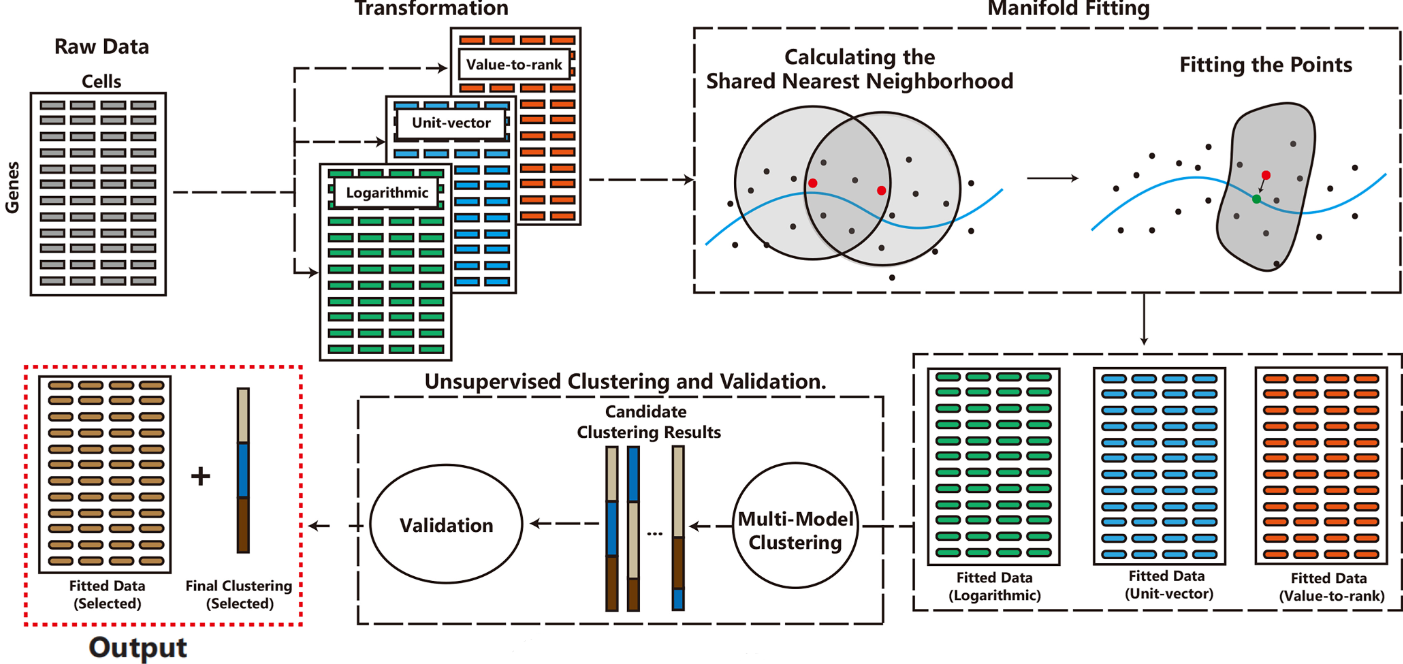}
    \caption{Overview of the scAMF framework for scRNA-seq data analysis. The schematic depicts the scAMF pipeline, which consists of three critical steps: raw data transformation, manifold fitting, and unsupervised clustering with subsequent validation. This methodology targets denoising, refinement of neighborhood structure, and more informative clustering by exploiting the latent manifold structure inherent in the data.}
    \label{fig:scAMF}
\end{figure}

\subsubsection*{Application in single-cell RNA sequencing data} 
Single-cell RNA sequencing (scRNA-seq) is critical to genomic research, providing detailed insights into cellular diversity and disease mechanisms. However, inherent noise from biological variability and technical factors complicates data analysis. Traditional methods struggle to accurately characterize cellular relationships due to this noise.

The scAMF framework, presented in Figure \ref{fig:scAMF} and proposed by \citet{scAMF}, is designed to denoise scRNA-seq data, recover more biologically meaningful neighborhood structure, and support more reliable clustering by fitting a low-dimensional manifold into the ambient gene expression data space. This approach reduces noise while preserving essential biological information. scAMF denoises scRNA-seq data by unfolding its distribution to reconstruct a smooth manifold that captures the underlying structure of the data with limited distortion of biologically relevant variation.

Central to the effectiveness of scAMF is its ability to reshape the spatial distribution of gene expression vectors so that cells of the same type lie closer together while different cell types are more clearly separated. This improved geometry supports more reliable clustering and clearer low-dimensional visualization. In the reported comparisons, scAMF was evaluated through clustering metrics such as ARI, NMI, and ACC, together with neighborhood- and visualization-oriented criteria such as neighborhood purity, intra/interclass distance, and silhouette index, and was found to perform favorably across diverse datasets. These features make scAMF a useful tool for single-cell analysis, with the potential to reveal cellular heterogeneity and rare cell populations more clearly. A related downstream development, CellScope \citep{li2025cellscope}, further highlighted the biological interpretability of this line of work by identifying disease-associated states in PBMC data from COVID-19 patients. In particular, it distinguished disease-related states within the monocyte--dendritic cell system and identified eight marker genes with progressive expression changes across healthy, moderate, and severe conditions, providing additional insight into antiviral immune responses.

\subsubsection*{Application in privacy-preserving analysis of sensitive data. }
In many modern applications, the data sets that exhibit low-dimensional manifold structure are also highly sensitive, as in genomics, biomedicine, and finance. This creates a tension between geometric learning and privacy protection: manifold-based methods seek to exploit local geometric structure, whereas formal privacy guarantees require limiting the influence of any individual record on the released output. Recent work has begun to address this tension by incorporating differential privacy into manifold-based denoising and geometric estimation. In particular, \citet{wu2025differential} study a private-reference, public-query setting in which a sensitive reference data set is used to denoise new noisy query points without exposing the underlying private data.

The key idea is to privatize local geometric summaries rather than the raw observations themselves. Their framework estimates local means and tangent-space information under calibrated $(\varepsilon,\delta)$-differential privacy, and then uses these privatized summaries to guide iterative denoising toward the latent manifold. From a statistical perspective, this leads to an explicit interplay among geometric bias, sampling variability, and privacy-induced perturbation. Theoretical analysis provides non-asymptotic utility guarantees that quantify how privacy noise affects geometric recovery, while simulations illustrate clear privacy--utility trade-offs under varying privacy budgets.

Beyond theory, this line of work also demonstrates that privacy-aware manifold methods can remain useful in realistic scientific settings. Case studies on UK Biobank biomarker data and single-cell RNA sequencing data show that important local geometric features, such as neighborhood structure and tangent information, can still be retained under moderate privacy budgets, allowing downstream analyses such as clustering and risk modeling to remain informative. These results suggest that manifold-based methodology can be extended in a principled way to regulated environments where direct access to individual-level data is restricted.

\section{Outlook}
In this review, we have traced the development of manifold fitting from early nonparametric ideas to mathematically grounded constructions and more practicable statistical methods, and we have highlighted several applications in modern data analysis. Compared with related techniques such as manifold embedding and manifold denoising, manifold fitting is distinguished by its explicit geometric target: it seeks to recover a smooth manifold object in the ambient space together with useful local and global geometric structure. As the literature continues to expand, it is natural to ask not only how current methods differ, but also what the next major problems for the field should be.

We close by highlighting several open problems and future directions that, in our view, arise naturally from the current state of manifold fitting.
\begin{itemize}[leftmargin=0.5cm]
    \item \textbf{Problem 1 (Flow-based fitting and the use of prior geometric structure).}\\
    Most current manifold fitting methods are essentially static: they estimate a single geometric object from an observed cloud of points. However, many important data sets are generated by trajectories, flows, or other forms of structured geometric evolution. This suggests a need for manifold fitting procedures that interact more directly with dynamical information, rather than treating the sample only as an unordered set. Relatedly, in some applications one may already possess partial geometric information, such as symmetry, reference geometry, boundary constraints, or known local directions, and it remains unclear how such prior structure should be incorporated in a principled way. More broadly, connections with geometry-aware low-dimensional representation and nested geometric modeling, such as \citet{su2025principal}, as well as recent principal-submanifold-based approaches to clustering and multiscale RNA correction \citep{wu2025rna}, suggest that future work may move beyond estimating a single fitted set toward richer structured descriptions of variation on or near manifolds.
    
    \item \textbf{Problem 2 (Statistical limits under curvature and unbounded noise).}\\
    Another central question concerns the statistical limits of manifold fitting under realistic geometric and stochastic regimes. Throughout this review, quantities such as smoothness, reach, and projection stability appear repeatedly, but much remains unknown about how these interact sharply with sampling scale, ambient dimension, and noise level. In particular, high curvature, near-self-intersection, and unbounded-noise settings raise basic questions about what geometric features remain identifiable and what rates are fundamentally achievable. A representative example is the limiting accuracy of local PCA and related tangent-space recovery procedures under unbounded noise, since these directly constrain downstream fitting accuracy. Clarifying such sharp lower and upper bounds would substantially deepen the theoretical foundation of manifold fitting.
    
    \item \textbf{Problem 3 (Geometric inference beyond set recovery).}\\
    A further direction is to move beyond estimating the manifold only as a set and instead estimate geometric quantities carried by the manifold. Tangent spaces, intrinsic dimension, second fundamental form, curvature-related objects, and more generally quadratic or second-order local structure are often the quantities most directly connected to scientific interpretation. Recent work has begun to explore this broader program of geometric inference beyond set recovery; examples include curvature-driven manifold fitting under unbounded isotropic noise \citep{li2026curvature} and the estimation of Riemannian quantities from noisy data via density derivatives \citep{chenliyao2026riemannian}. These developments suggest that second-order structure may become an increasingly important target in its own right, linking manifold fitting more directly to questions of curvature, local approximation, and geometric interpretability.
    
    \item \textbf{Problem 4 (Manifold fitting beyond Euclidean ambient spaces and transformed representations).}\\
    Most current manifold fitting methods are formulated for data embedded in Euclidean ambient spaces. In practice, however, many modern analyses are carried out only after substantial feature transformation, representation learning, or embedding into another space, and some data objects are intrinsically non-Euclidean from the outset. This raises basic questions about how manifold fitting should interact with embedding transforms, how geometric meaning can be preserved after such transformations, and how low-dimensional manifold ideas can be extended to more general ambient spaces. The challenge is not simply technical generalization, but also interpretability: one would like fitted geometric structure in the transformed or non-Euclidean setting to remain meaningful for the original scientific problem.
    
    \item \textbf{Problem 5 (Manifold-constrained generative modeling and diffusion on unknown manifold spaces).}\\
    Recent advances in score-based generative models and diffusion processes have shown strong performance for high-dimensional data, which are often believed to concentrate near low-dimensional manifolds. This suggests a natural connection with manifold fitting: rather than evolving isotropically in the ambient space, generative mechanisms could in principle be constrained by an estimated manifold and its local geometry. In particular, one may ask whether diffusion should be defined intrinsically on the manifold, with stochastic evolution adapted to tangent directions rather than unconstrained ambient perturbations. Recent theoretical work \citet{pmlr-v291-potaptchik25a} suggests that, under the manifold hypothesis, the convergence behavior of diffusion models may depend primarily on the intrinsic rather than ambient dimension. However, such analyses typically assume that the underlying manifold is known or that the geometric setting is idealized. In statistical applications, the manifold itself must first be estimated from noisy data, after which the diffusion process must be constructed on the fitted manifold. Understanding how manifold estimation error propagates into diffusion dynamics, generative accuracy, and intrinsic-dimension-dependent efficiency remains largely open. More broadly, this points toward a unified framework combining manifold fitting, stochastic processes on manifolds, and modern generative modeling, in which geometric structure plays a central role in both theory and computation.
    
    \item \textbf{Problem 6 (Stratified and multi-manifold structure).}\\
    Another emerging direction concerns manifold fitting under stratified or multi-manifold structure, where data arise from a union of multiple manifolds, possibly of different intrinsic dimensions and with nontrivial intersections. Such settings naturally arise in many applications, including heterogeneous populations, branching dynamical systems, and multi-state biological processes, where a single manifold assumption may fail to capture the underlying structure. Recent theoretical work, such as \citep{aamari2024theory}, has begun to study the statistical limits of such problems, establishing minimax rates and identifiability conditions for recovering stratified geometric structure. However, these results are largely of a theoretical nature, and it remains unclear how to translate them into practically scalable and robust algorithms, particularly in high-dimensional or noisy settings. Moreover, standard local methods may break down near regions of overlap, where multiple tangent structures coexist. From a statistical perspective, this setting introduces an additional layer of complexity through the interaction between geometric estimation and latent structure assignment. Overall, stratified manifold models represent a promising but still largely undeveloped extension of current manifold fitting methodology.

    \item \textbf{Problem 7 (Global parametrization and coordinate representations for manifold fitting).}\\
    A largely unexplored direction in manifold fitting concerns the systematic construction and statistical exploitation of explicit global coordinate systems. Most existing fitting procedures either operate intrinsically on the manifold or embed it into a high-dimensional ambient Euclidean space, without exploiting the possibility of nearly global parametrizations that cover almost all of the manifold while excluding only a negligible singular set. Recent work of Li and Yao \citep{li2026approx} develops a systematic theory of \emph{approximate parametrization} for smooth manifolds, providing explicit diffeomorphisms from open dense subsets of classical manifolds---including Lie groups, Stiefel manifolds, Grassmannians, flag manifolds, and symmetric spaces---onto Euclidean domains, together with closed-form inverse maps. The excluded \emph{parametrization singular set} is shown to be a null set whose codimension is controlled by the first non-trivial Betti number of the manifold, and this bound is sharp across all classical examples. This raises natural and largely open questions for manifold fitting: whether proximity to the singular set induces bias or variance inflation in geometric estimators, how the non-trivial Jacobian weight induced by the parametrization map affects minimax estimation rates, and how coordinate-based fitting methods can be designed to remain statistically well-behaved near the singular boundary. More broadly, understanding how approximate parametrizations interact with curvature estimation, goodness-of-fit testing, and the interpretability of fitted geometric structure---particularly when the ambient space is itself non-Euclidean or obtained through a representation transform---represents a promising direction for future work.
    \item \textbf{Problem 8 (Privacy-aware geometric inference).}\\
    As manifold-based methods move toward increasingly sensitive domains, an important open direction is how to preserve useful geometric structure while enforcing rigorous privacy protection. Recent work has shown that differential privacy can be incorporated into manifold denoising and related geometric estimation problems by privatizing local geometric summaries such as means and tangent information. A central challenge for future work is to understand more sharply how privacy constraints affect geometric recovery, both statistically and computationally, and how such methods can be extended from denoising to broader forms of manifold fitting and geometric inference.
\end{itemize}

Taken together, these open problems suggest that manifold fitting is evolving from a method for geometric reconstruction toward a broader framework for structured geometric inference. Future progress will likely depend on combining geometric insight, statistical optimality, computational tractability, and application-specific considerations in a more unified way. We hope that the present review provides a useful foundation for these developments.

\subsection*{Disclosure Statement}
The authors have no conflicts of interest to declare.

\subsection*{Acknowledgments}
Zhigang Yao has received support from the Singapore Ministry of Education through the Tier 2 grants (A-0008520-00-00 and A-8001562-00-00) and Tier 1 grants (A-8004146-00-00 and A-8002931-00-00) at the National University of Singapore. Jiaji Su is a postdoctoral researcher supported by the grant A-8001562-00-00. Zhigang Yao also acknowledges the support of the Center for Mathematical Sciences and Applications (CMSA) at Harvard University since 2022.

This work originated from the Harvard Conference on Geometry and Statistics, hosted by CMSA and held from February 27 to March 1, 2023. A substantial portion of the survey work, including the two manifold-based applications, was completed during Zhigang Yao’s visit to Tsinghua University in 2023. The project was completed during Zhigang Yao's recent visit to the Shanghai Institute of Mathematics and Interdisciplinary Sciences (SIMIS) in 2024.
The authors thank Xiao-Li Meng, the founding editor of {\it Harvard Data Science Review}, for the invitation to contribute to this special theme.


\appendix


\bibliographystyle{plainnat}
\bibliography{references}

\end{document}